%% file: main.tex
\documentclass[11pt]{article}
\usepackage[a4paper,margin=1in]{geometry}
\usepackage[T1]{fontenc}\usepackage[utf8]{inputenc}
\usepackage{lmodern,microtype,amsmath,amssymb,booktabs,tabularx,longtable,array,fancyhdr}
\usepackage{subcaption}
\usepackage{multirow}
\usepackage{enumitem}
\usepackage{needspace}
\usepackage{amsmath}
\usepackage{pgfplots}
\pgfplotsset{compat=1.18}
\usepgfplotslibrary{groupplots}
\usepackage{tikz}
\usepackage{float}
\definecolor{colpick}{HTML}{D55E00}
\usetikzlibrary{arrows.meta,backgrounds,shapes.geometric,positioning}
\definecolor{colmove}{HTML}{0072B2}
\definecolor{colshifttwo}{HTML}{14456B}
\definecolor{coltarget}{HTML}{D55E00}
\definecolor{colcomm}{HTML}{009E73}
\definecolor{colconvex}{HTML}{0072B2}
\definecolor{collouvain}{HTML}{E69F00}
\definecolor{colwalk}{HTML}{009E73}
\definecolor{colhoptwo}{HTML}{D55E00}
\definecolor{colhopone}{HTML}{CC79A7}
\definecolor{colorig}{HTML}{BFD8E8}
\definecolor{colmod}{HTML}{E8C6DC}
\definecolor{colfused}{HTML}{E8A33D}
\definecolor{colrot}{HTML}{C0812A}
\definecolor{colshift}{HTML}{2E7FB8}
\definecolor{colexp}{HTML}{CC79A7}   
\definecolor{colans}{HTML}{FBE3D1}
\definecolor{colwin}{HTML}{0072B2}
\definecolor{colctrl}{HTML}{E69F00}
\definecolor{colplain}{HTML}{009E73}
\usepackage[hidelinks]{hyperref}
\newcolumntype{Y}{>{\raggedright\arraybackslash}X}
\newcommand{\code}[1]{\texttt{#1}}
\newcommand{\fixed}{\textit{fixed}}

\title{\textbf{Graph-free Retrieval with Graph-Modified Embeddings}}
\author{
	Federica Santi\thanks{Correspondence: \href{mailto:federica@topoteretes.com}{\texttt{federica@topoteretes.com}}.} \quad
	Vasilije Markovi\'{c} \quad
	Daniel Scott \quad
	Lazar Obradovi\'{c}\\[0.3em]
	\normalsize Cognee Inc.
}
\date{}
\begin{document}
\maketitle\vspace{-2em}
\input{src/abstract}
\input{src/introduction}
\input{src/related_work}
\input{src/setup}
\input{src/experiments}
\input{src/retrieval_analysis}
\input{src/conclusions}
\input{src/credit}

\newpage

\input{src/bibliography}
\newpage
\appendix
\input{src/appendix_overview}
\input{src/appendix_data}

\input{src/appendix_retrofitting}

\input{src/appendix_retrieval}

\input{src/appendix_per_family}
\end{document}

%% file: src/abstract.tex
\begin{abstract}
	\textit{Dense vector retrieval can miss evidence connected through intermediate entities, while
		direct graph retrieval requires access to the graph at query time. We compare five offline
		graph-based embedding transformations and strategies for using the resulting graph-modified
		embeddings alongside the originals, without query-time graph access. On a synthetic
		professional-network benchmark with complete supporting-evidence sets, direct replacement
		of the original embeddings reduces evidence recovery. Multi-stage strategies improve recovery
		by using graph-modified embeddings to guide the search while retaining the originals for
		semantic relevance. On questions requiring linked facts or comparisons among candidates,
		the strongest configuration improves recovery by up to 25.9\,\% relative to plain vector retrieval.
		It also matches baseline recovery with 40\,\% fewer returned nodes. Controls confirm that graph
		information contributes beyond multi-stage retrieval alone. Across the tested methods,
		retrieval performance depends more on how graph-modified embeddings are used than on how
		they are produced.
	}
\end{abstract}

%% file: src/introduction.tex
\section{Introduction}
Retrieval is a central component of retrieval-augmented generation (RAG), where external evidence is retrieved to support large language model (LLM) outputs \cite{lewis2020rag}. It also supports LLM agents and long-term memory systems by identifying relevant documents, passages, or stored facts. Dense retrieval embeds queries and passages in a shared vector space and ranks candidates by vector similarity \cite{karpukhin2020dpr}.
This approach is effective when supporting
evidence is semantically close to the query, but may miss relevant
passages connected through intermediate entities or multi-step
relations.
Knowledge graphs represent these connections explicitly, but using them for retrieval introduces two challenges. Traversing or reranking the graph for every query increases query-time complexity, while modifying embeddings according to graph structure may distort the semantic relationships captured by the original embeddings. This leads to the central question of this work: \emph{can graph structure be incorporated offline into graph-modified embeddings that improve retrieval without requiring graph access at query time or replacing the original embeddings?}

Prior work has either incorporated relational knowledge into embeddings~\cite{faruqui,mrksic,lengerich}, used graphs directly at query time~\cite{graftnet,pullnet}, or refined retrieval through textual feedback alone~\cite{nprf,colbertprf,yuprf}. Our work connects these directions by examining both how graph-modified embeddings should be constructed and how they should be used in multi-stage retrieval.

We construct graph-modified embeddings offline using several transformations, including local neighbourhood aggregation, random-walk endpoints, community centroids~\cite{blondel}, and convex neighbour pull. We then compare retrieval strategies ranging from replacing the original embeddings with graph-modified embeddings to multi-stage strategies that combine both. At query time, every strategy uses only cosine-similarity searches over stored embeddings; the graph is neither traversed nor queried.

Our central hypothesis is that graph-modified embeddings are more useful for guiding the expansion of retrieval than for measuring relevance to the original query. The best-performing strategies therefore assign distinct roles to the two sets of embeddings. An initial search over the original embeddings retrieves semantically relevant results (seeds). Their graph-modified embeddings are then aggregated to guide the selection of candidates that are both relevant to the query and related to the seeds through the graph. Finally, the original embeddings of the selected candidates are used to reformulate the query for another search over the original embeddings~\cite{nprf,colbertprf,yuprf}. We refer to this use of graph-modified embeddings to guide retrieval as \emph{graph conditioning}.

To evaluate these strategies, we introduce a synthetic professional-network benchmark in which the corpus, graph, questions, answers, and supporting evidence are generated from the same underlying facts. The benchmark includes single-fact, relational and collection, and extremisation questions, together with exact gold answers and complete gold-evidence sets containing the intermediate and competing records required to establish each answer~\cite{cattaneo}.

We select configurations on a development corpus and report results on separately generated validation data (Section~\ref{sec:split}). All methods use the same original embeddings, cosine similarity, and the same 40-node retrieval limit. Multi-stage controls without graph conditioning allow us to distinguish improvements produced by graph-modified embeddings from those produced by multi-stage retrieval alone.

None of the graph-modified representations improves direct retrieval over the original embeddings in our experiments. The original embeddings recover $73.1\,\%$ of the gold evidence, whereas the five transformations recover between $52\,\%$ and $72.9\,\%$. In contrast, multi-stage retrieval improves performance across all five transformations, and the choice of retrieval strategy has a greater effect than the transformation used to construct the graph-modified embeddings. The strongest configuration improves gold-evidence recovery relative to plain vector retrieval by $25.9\,\%$ on extremisation questions and by $10.0\,\%$ on relational and collection questions ($19.2\,\%$ across the two groups). It also matches the gold-evidence recovery achieved by plain vector retrieval with 40 nodes after returning only 24, corresponding to $40\,\%$ fewer returned nodes.
\paragraph{Contributions.}
\begin{itemize}
	\item We provide a controlled comparison of multiple graph-based
	embedding transformations and retrieval strategies under a common
	cosine-similarity framework, without requiring graph access at query time.

	\item We introduce a synthetic benchmark with questions of different
	relational complexity, exact gold answers, and complete
	question-specific supporting evidence.

	\item We show that retrieval performance depends more on how
	graph-modified embeddings are used than on the particular
	transformation that produces them, and that they are most effective as
	intermediate signals within a multi-stage retriever.
	\item We introduce a retrieval strategy in which graph-modified embeddings guide exploration while the original embeddings preserve semantic alignment with the query, and we quantify its performance.

\end{itemize}

%% file: src/related_work.tex
\section{Related Work}

Our work connects research on graph-modified embeddings,
feedback-based and multi-hop retrieval, and graph-based retrieval.
We also draw on evidence-level evaluation to assess whether these
approaches recover the passages needed to answer a question.

\paragraph{Encoding graph structure into embeddings.}
Retrofitting adapts pretrained embeddings using external relational
knowledge. Faruqui et al.~\cite{faruqui} introduced an
objective that balances fidelity to the original embeddings with
attraction between connected concepts. Counter-fitting adds
attraction and repulsion constraints~\cite{mrksic}, while Functional
Retrofitting introduces relation-specific transformations for
heterogeneous knowledge graphs~\cite{lengerich}. Retrofitting has
also been applied beyond word embeddings, including multilingual
sentence representations aligned through Abstract Meaning
Representation~\cite{cai} and sense representations specialised
using WordNet relations~\cite{mizuki}. Related approaches use entity
descriptions and language models to construct representations for
knowledge-graph completion, incorporating structural information
through different training objectives or architectures
\cite{xie,wang,li,qiu,papaluca}.

Another line of work derives node representations from graph
connectivity. DeepWalk learns embeddings from contexts generated
by truncated random walks~\cite{perozzi2014deepwalk}, while node2vec
uses biased walks to capture different network neighbourhoods
\cite{grover2016node2vec}. These methods are conceptually related
to our random-walk transformation, which instead averages existing
textual embeddings at walk endpoints without learning new
representations from walk sequences. Simple Graph Convolution
separates graph-based feature propagation from a learned
classifier~\cite{wu2019sgc}, providing a related perspective on
our neighbourhood-based aggregation of existing embeddings.

We draw on these ideas to study graph-modified embeddings
for evidence retrieval, rather than intrinsic similarity, node
classification, or knowledge-graph completion. In particular, we
compare their use as direct search representations with their use
as intermediate signals in a multi-stage retriever.
\paragraph{Feedback-based and multi-hop dense retrieval.}
Pseudo-relevance feedback refines a query using the results of an initial retrieval stage and then performs a second search. NPRF (neural pseudo-relevance feedback) brought the idea into learned ranking, feeding the top-ranked documents of a first pass into a neural ranking model~\cite{nprf}. More recent approaches apply the same principle to dense retrieval, where the query is a vector rather than a set of terms, and differ in what that vector is allowed to be. ColBERT-PRF keeps a multi-vector query and extends it, extracting representative embeddings from the pseudo-relevant documents and adding them to the query representation before re-retrieval~\cite{colbertprf}. ANCE-PRF keeps a single query vector but learns an encoder that consumes the initially retrieved passages, leaving the document index unchanged~\cite{yuprf}. Multi-hop retrieval extends this sequential approach to questions whose
supporting evidence cannot be identified from the original query alone.
Multi-Hop Dense Retrieval conditions successive searches on previously
retrieved passages to recover chains of supporting evidence
\cite{xiong2021multihop}. Other approaches use generated text to guide
subsequent retrieval. IRCoT interleaves retrieval with intermediate
chain-of-thought reasoning steps \cite{trivedi2023ircot}, while Iter-RetGen
alternates retrieval and generation, using each generated output to
inform the next retrieval iteration \cite{shao2023iterretgen}.

Our method uses offline graph-modified embeddings to guide
intermediate retrieval, followed by a Rocchio-style update that blends the query
with the mean embedding of retrieved nodes in the original space \cite{rocchio1971relevance}, without query-time
graph access or LLM generation. Controls without graph conditioning (Section~\ref{sec:graph-conditioning})
isolate the contribution of graph conditioning.

\paragraph{Graph-based retrieval.}
Graph-based systems use explicit relations to find evidence that
semantic similarity may miss, such as education records linked
to a company through its employees. GRAFT-Net reasons over
question-specific subgraphs of facts and text~\cite{graftnet};
PullNet expands them through iterative retrieval~\cite{pullnet};
and HippoRAG propagates relevance through entity connections
using Personalized PageRank~\cite{gutierrez2024hipporag}.
These systems use the graph during query processing, whereas
we encode its relations into embeddings offline and retrieve
evidence through stored vectors alone.

GraphRAG uses community summaries to answer global questions
about a corpus~\cite{edge2024graphrag}. Although related to our
use of graph communities, it produces text for corpus-level
synthesis, whereas we produce node-level vectors for retrieving
question-specific evidence under a fixed budget.
\paragraph{Evidence-level evaluation.}
Multi-hop question-answering benchmarks support evaluation beyond
final-answer accuracy by identifying the evidence needed for
reasoning. HotpotQA provides sentence-level supporting facts
alongside questions and answers, including both bridge and
comparison questions \cite{yang2018hotpotqa}. MuSiQue constructs
multi-hop questions by composing single-hop questions whose
reasoning steps depend on one another, explicitly addressing
shortcuts that allow apparently multi-hop questions to be answered
without combining the required evidence \cite{trivedi2022musique}.
Ground-truth-subgraph benchmarks such as SynthKGQA and GTSQA
further support the evaluation of graph retrieval independently
of answer generation \cite{cattaneo}.
We adopt this evidence-centred perspective: our synthetic corpus
and questions are generated from the same structured facts,
enabling us to specify question-specific gold-evidence sets and
measure their recovery directly. The benchmark includes relational and
collection questions and extremisation questions, allowing us to examine
retrieval of both indirectly connected evidence and the competing
records required to establish an extremum.

%% file: src/setup.tex
	\section{Experimental Setup}
	We evaluate how graph-based embedding transformations and retrieval strategies affect
	evidence recovery on a synthetic professional-network benchmark.

	\subsection{Dataset and Knowledge Graph}\label{sec:split}
	We use separate development and validation corpora, generated by the same pipeline
	and containing synthetic resumes and organisational personnel rosters.

	Each corpus is represented as a tripartite graph with three node types: documents, chunks (textual
	evidence from those documents), and entities (people and organisations). Three relation
	types connect them: \texttt{part\_of} links chunks to documents, \texttt{about} links
	documents to their primary entities, and \texttt{mentions} links chunks to entities
	referenced in their text. The schema admits only these three edge patterns.
	Each corpus contains approximately 1,200 chunks, 370 entities, 230 documents, and 3,700 edges.
	The development and main validation corpora are node-disjoint: none of the 1,759 node identifiers in the validation
	corpus occurs in the development corpus.
	We use Cognee\footnote{An open-source AI memory framework combining knowledge graphs and
		vector search: \url{https://github.com/topoteretes/cognee}.} to ingest and store the generated
	graph and manage its associated node embeddings.

	We select configurations on the development corpus and report results on the validation
	set of $977$ questions, including $90$ generated on a supplementary corpus from the same pipeline.
	Appendix~\ref{app:data} provides generation details, exact validation statistics, and
	the schema with an example (Figure~\ref{fig:schema}).

	\subsection{Questions and Gold Evidence}
	We generate questions and ground-truth (gold) answers from the structured facts underlying the dataset.
	Gold evidence includes every chunk required to establish an answer. The benchmark contains
	11 question families, detailed in Appendix~\ref{app:data}. We group them into three categories:
	\begin{description}[leftmargin=0pt, labelsep=0.6em, itemsep=3pt]
		\item[Single-fact questions:] These ask for attributes directly associated with
		entities mentioned in the query and generally require a single evidence chunk or an
		immediate relation. For example, ``What skills did a person use at a given employer?''
		can be answered using a job chunk that mentions the person, employer, and skills.
		\item[Relational and collection questions:] These ask for attributes or entities
		connected to a named entity through intermediate entities. For example, ``Which
		universities were attended by the employees of an organisation?'' requires evidence
		linking the organisation to its employees and those employees to their universities.
		The target chunks may lie two or three relations from the named entity and share
		little surface content with the query.
		\item[Extremisation questions:] These compare candidates to identify a maximum
		and may then request another attribute of the selected candidate. Establishing the
		maximum requires the relevant records of all competing candidates. For example,
		``What certifications does the most experienced worker at a given company hold?''
		requires experience records for the comparison and the winner's certification evidence,
		which may be semantically unrelated to the comparison criterion.
	\end{description}

	\subsection{Graph-Based Embedding Transformations}\label{emb-strategy}

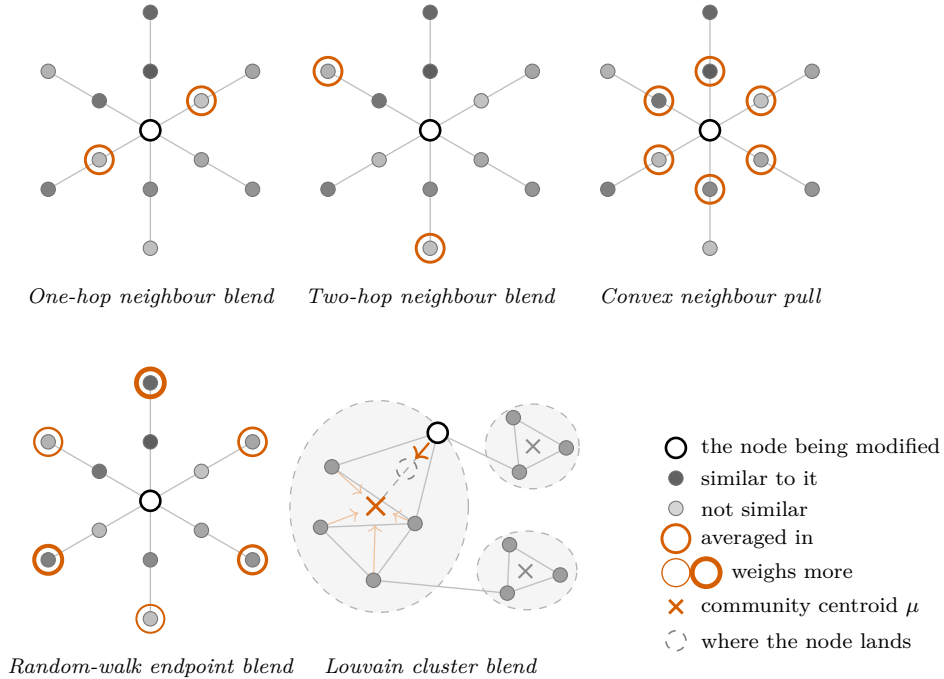
\begin{figure}[H]
		\centering
		\begin{tikzpicture}[
			x=1cm, y=1cm, font=\scriptsize,
			nd/.style={circle, draw=black!55, minimum size=5.4pt, inner sep=0pt},
			focal/.style={circle, draw=black, line width=1pt, fill=white, minimum size=7.5pt,
				inner sep=0pt},
			pick/.style={circle, draw=colpick, line width=1.1pt, minimum size=10.5pt, inner sep=0pt},
			edge/.style={draw=black!25, line width=0.5pt},
			title/.style={font=\scriptsize\itshape, anchor=north},
			sub/.style={font=\scriptsize, anchor=north, text=black!60},
			comm/.style={draw=black!30, dashed, line width=0.5pt, fill=black!4},
			pull/.style={->, draw=colpick, line width=0.9pt},
			pullfaint/.style={->, draw=colpick!40, line width=0.45pt},
			ghost/.style={circle, draw=black!50, dashed, line width=0.6pt,
				minimum size=7.5pt, inner sep=0pt},
			]
			\begin{scope}[shift={(0,0)}]
				\draw[edge] (0,0) -- (0.0,0.78);
				\draw[edge] (0,0) -- (-0.675,0.39);
				\draw[edge] (0,0) -- (-0.675,-0.39);
				\draw[edge] (0,0) -- (-0.0,-0.78);
				\draw[edge] (0,0) -- (0.675,-0.39);
				\draw[edge] (0,0) -- (0.675,0.39);
				\draw[edge] (0.0,0.78) -- (0.0,1.56);
				\draw[edge] (-0.675,0.39) -- (-1.351,0.78);
				\draw[edge] (-0.675,-0.39) -- (-1.351,-0.78);
				\draw[edge] (-0.0,-0.78) -- (-0.0,-1.56);
				\draw[edge] (0.675,-0.39) -- (1.351,-0.78);
				\draw[edge] (0.675,0.39) -- (1.351,0.78);
				\node[nd, fill=black!59] at (0.0,1.56) {};
				\node[nd, fill=black!30] at (-1.351,0.78) {};
				\node[nd, fill=black!50] at (-1.351,-0.78) {};
				\node[nd, fill=black!25] at (-0.0,-1.56) {};
				\node[nd, fill=black!42] at (1.351,-0.78) {};
				\node[nd, fill=black!35] at (1.351,0.78) {};
				\node[nd, fill=black!63] at (0.0,0.78) {};
				\node[nd, fill=black!54] at (-0.675,0.39) {};
				\node[nd, fill=black!27] at (-0.675,-0.39) {};
				\node[nd, fill=black!46] at (-0.0,-0.78) {};
				\node[nd, fill=black!34] at (0.675,-0.39) {};
				\node[nd, fill=black!24] at (0.675,0.39) {};
				\node[focal] at (0,0) {};
				\node[pick] at (0.675,0.39) {};
				\node[pick] at (-0.675,-0.39) {};
				\node[title] at (0,-1.95) {One-hop neighbour blend};
			\end{scope}
			\begin{scope}[shift={(3.7,0)}]
				\draw[edge] (0,0) -- (0.0,0.78);
				\draw[edge] (0,0) -- (-0.675,0.39);
				\draw[edge] (0,0) -- (-0.675,-0.39);
				\draw[edge] (0,0) -- (-0.0,-0.78);
				\draw[edge] (0,0) -- (0.675,-0.39);
				\draw[edge] (0,0) -- (0.675,0.39);
				\draw[edge] (0.0,0.78) -- (0.0,1.56);
				\draw[edge] (-0.675,0.39) -- (-1.351,0.78);
				\draw[edge] (-0.675,-0.39) -- (-1.351,-0.78);
				\draw[edge] (-0.0,-0.78) -- (-0.0,-1.56);
				\draw[edge] (0.675,-0.39) -- (1.351,-0.78);
				\draw[edge] (0.675,0.39) -- (1.351,0.78);
				\node[nd, fill=black!59] at (0.0,1.56) {};
				\node[nd, fill=black!30] at (-1.351,0.78) {};
				\node[nd, fill=black!50] at (-1.351,-0.78) {};
				\node[nd, fill=black!25] at (-0.0,-1.56) {};
				\node[nd, fill=black!42] at (1.351,-0.78) {};
				\node[nd, fill=black!35] at (1.351,0.78) {};
				\node[nd, fill=black!63] at (0.0,0.78) {};
				\node[nd, fill=black!54] at (-0.675,0.39) {};
				\node[nd, fill=black!27] at (-0.675,-0.39) {};
				\node[nd, fill=black!46] at (-0.0,-0.78) {};
				\node[nd, fill=black!34] at (0.675,-0.39) {};
				\node[nd, fill=black!24] at (0.675,0.39) {};
				\node[focal] at (0,0) {};
				\node[pick] at (-0.0,-1.56) {};
				\node[pick] at (-1.351,0.78) {};
				\node[title] at (0,-1.95) {Two-hop neighbour blend};
			\end{scope}
			\begin{scope}[shift={(7.4,0)}]
				\draw[edge] (0,0) -- (0.0,0.78);
				\draw[edge] (0,0) -- (-0.675,0.39);
				\draw[edge] (0,0) -- (-0.675,-0.39);
				\draw[edge] (0,0) -- (-0.0,-0.78);
				\draw[edge] (0,0) -- (0.675,-0.39);
				\draw[edge] (0,0) -- (0.675,0.39);
				\draw[edge] (0.0,0.78) -- (0.0,1.56);
				\draw[edge] (-0.675,0.39) -- (-1.351,0.78);
				\draw[edge] (-0.675,-0.39) -- (-1.351,-0.78);
				\draw[edge] (-0.0,-0.78) -- (-0.0,-1.56);
				\draw[edge] (0.675,-0.39) -- (1.351,-0.78);
				\draw[edge] (0.675,0.39) -- (1.351,0.78);
				\node[nd, fill=black!59] at (0.0,1.56) {};
				\node[nd, fill=black!30] at (-1.351,0.78) {};
				\node[nd, fill=black!50] at (-1.351,-0.78) {};
				\node[nd, fill=black!25] at (-0.0,-1.56) {};
				\node[nd, fill=black!42] at (1.351,-0.78) {};
				\node[nd, fill=black!35] at (1.351,0.78) {};
				\node[nd, fill=black!63] at (0.0,0.78) {};
				\node[nd, fill=black!54] at (-0.675,0.39) {};
				\node[nd, fill=black!27] at (-0.675,-0.39) {};
				\node[nd, fill=black!46] at (-0.0,-0.78) {};
				\node[nd, fill=black!34] at (0.675,-0.39) {};
				\node[nd, fill=black!24] at (0.675,0.39) {};
				\node[focal] at (0,0) {};
				\node[pick] at (0.0,0.78) {};
				\node[pick] at (-0.675,0.39) {};
				\node[pick] at (-0.675,-0.39) {};
				\node[pick] at (-0.0,-0.78) {};
				\node[pick] at (0.675,-0.39) {};
				\node[pick] at (0.675,0.39) {};
				\node[title] at (0,-1.95) {Convex neighbour pull};
			\end{scope}
			\begin{scope}[shift={(0,-4.9)}]
				\draw[edge] (0,0) -- (0.0,0.78);
				\draw[edge] (0,0) -- (-0.675,0.39);
				\draw[edge] (0,0) -- (-0.675,-0.39);
				\draw[edge] (0,0) -- (-0.0,-0.78);
				\draw[edge] (0,0) -- (0.675,-0.39);
				\draw[edge] (0,0) -- (0.675,0.39);
				\draw[edge] (0.0,0.78) -- (0.0,1.56);
				\draw[edge] (-0.675,0.39) -- (-1.351,0.78);
				\draw[edge] (-0.675,-0.39) -- (-1.351,-0.78);
				\draw[edge] (-0.0,-0.78) -- (-0.0,-1.56);
				\draw[edge] (0.675,-0.39) -- (1.351,-0.78);
				\draw[edge] (0.675,0.39) -- (1.351,0.78);
				\node[nd, fill=black!59] at (0.0,1.56) {};
				\node[nd, fill=black!30] at (-1.351,0.78) {};
				\node[nd, fill=black!50] at (-1.351,-0.78) {};
				\node[nd, fill=black!25] at (-0.0,-1.56) {};
				\node[nd, fill=black!42] at (1.351,-0.78) {};
				\node[nd, fill=black!35] at (1.351,0.78) {};
				\node[nd, fill=black!63] at (0.0,0.78) {};
				\node[nd, fill=black!54] at (-0.675,0.39) {};
				\node[nd, fill=black!27] at (-0.675,-0.39) {};
				\node[nd, fill=black!46] at (-0.0,-0.78) {};
				\node[nd, fill=black!34] at (0.675,-0.39) {};
				\node[nd, fill=black!24] at (0.675,0.39) {};
				\node[focal] at (0,0) {};
				\node[pick, line width=1.80pt] at (0.0,1.56) {};
				\node[pick, line width=0.89pt] at (-1.351,0.78) {};
				\node[pick, line width=1.52pt] at (-1.351,-0.78) {};
				\node[pick, line width=0.75pt] at (-0.0,-1.56) {};
				\node[pick, line width=1.24pt] at (1.351,-0.78) {};
				\node[pick, line width=1.03pt] at (1.351,0.78) {};
				\node[title] at (0,-1.95) {Random-walk endpoint blend};
			\end{scope}
			\begin{scope}[shift={(3.7,-4.9)}]
				\draw[comm] (-0.675,-0.075) ellipse (1.18 and 1.40);
				\draw[comm] ( 1.35, 0.72)   ellipse (0.62 and 0.56);
				\draw[comm] ( 1.28,-0.92)   ellipse (0.66 and 0.52);
				\draw[edge] ( 0.10, 0.90) -- (-0.20,-0.30);
				\draw[edge] ( 0.10, 0.90) -- (-1.30, 0.45);
				\draw[edge] (-1.30, 0.45) -- (-0.20,-0.30);
				\draw[edge] (-0.20,-0.30) -- (-1.45,-0.35);
				\draw[edge] (-1.45,-0.35) -- (-0.75,-1.05);
				\draw[edge] (-0.75,-1.05) -- (-0.20,-0.30);
				\draw[edge] ( 1.10, 1.10) -- ( 1.80, 0.70);
				\draw[edge] ( 1.80, 0.70) -- ( 1.18, 0.38);
				\draw[edge] ( 1.18, 0.38) -- ( 1.10, 1.10);
				\draw[edge] ( 1.05,-0.58) -- ( 1.72,-0.98);
				\draw[edge] ( 1.72,-0.98) -- ( 1.02,-1.22);
				\draw[edge] ( 1.02,-1.22) -- ( 1.05,-0.58);
				\draw[edge] ( 0.10, 0.90) -- ( 1.18, 0.38);
				\draw[edge] (-0.75,-1.05) -- ( 1.02,-1.22);
				\draw[pullfaint, shorten >=7pt] (-1.30, 0.45) -- (-0.72,-0.07);
				\draw[pullfaint, shorten >=7pt] (-1.45,-0.35) -- (-0.72,-0.07);
				\draw[pullfaint, shorten >=7pt] (-0.75,-1.05) -- (-0.72,-0.07);
				\draw[pullfaint, shorten >=7pt] (-0.20,-0.30) -- (-0.72,-0.07);
				\draw[draw=black!35, dashed, line width=0.5pt, shorten <=5pt, shorten >=4pt]
					( 0.10, 0.90) -- (-0.72,-0.07);
				\node[nd, fill=black!38] at (-1.30, 0.45) {};
				\node[nd, fill=black!38] at (-1.45,-0.35) {};
				\node[nd, fill=black!38] at (-0.75,-1.05) {};
				\node[nd, fill=black!38] at (-0.20,-0.30) {};
				\node[nd, fill=black!38] at ( 1.10, 1.10) {};
				\node[nd, fill=black!38] at ( 1.80, 0.70) {};
				\node[nd, fill=black!38] at ( 1.18, 0.38) {};
				\node[nd, fill=black!38] at ( 1.05,-0.58) {};
				\node[nd, fill=black!38] at ( 1.72,-0.98) {};
				\node[nd, fill=black!38] at ( 1.02,-1.22) {};
				\draw[colpick, line width=1.1pt] (-0.84,-0.19) -- (-0.60, 0.05);
				\draw[colpick, line width=1.1pt] (-0.84, 0.05) -- (-0.60,-0.19);
				\draw[black!45, line width=0.8pt] ( 1.27, 0.63) -- ( 1.45, 0.81);
				\draw[black!45, line width=0.8pt] ( 1.27, 0.81) -- ( 1.45, 0.63);
				\draw[black!45, line width=0.8pt] ( 1.17,-1.02) -- ( 1.35,-0.84);
				\draw[black!45, line width=0.8pt] ( 1.17,-0.84) -- ( 1.35,-1.02);
				\node[ghost] at (-0.31, 0.42) {};
				\node[focal] at ( 0.10, 0.90) {};
				\draw[pull, shorten <=5pt, shorten >=5pt] (0.10,0.90) -- (-0.31,0.42);
				\node[title] at (0,-1.95) {Louvain cluster blend};
			\end{scope}
			\begin{scope}[shift={(6.95,-4.9)}]
				\node[focal] at (0,0.7) {};   \node[anchor=west] at (0.2,0.7) {the node being modified};
				\node[nd, fill=black!62] at (0,0.3) {}; \node[anchor=west] at (0.2,0.3) {similar to it};
				\node[nd, fill=black!16] at (0,-0.1) {}; \node[anchor=west] at (0.2,-0.1) {not similar};
				\node[pick] at (0,-0.5) {};   \node[anchor=west] at (0.2,-0.5) {averaged in};
				\node[pick, line width=0.5pt] at (0,-0.95) {};
				\node[pick, line width=1.8pt] at (0.4,-0.95) {};
				\node[anchor=west] at (0.6,-0.95) {weighs more};
				\draw[colpick, line width=1.1pt] (-0.09,-1.31) -- (0.09,-1.49);
				\draw[colpick, line width=1.1pt] (-0.09,-1.49) -- (0.09,-1.31);
				\node[anchor=west] at (0.2,-1.40) {community centroid $\mu$};
				\node[ghost] at (0,-1.85) {};
				\node[anchor=west] at (0.2,-1.85) {where the node lands};
			\end{scope}
		\end{tikzpicture}
		\caption{Graph information used by the five transformations. Darker nodes are more
		similar to the source node; orange outlines mark embeddings included in an average,
		with thicker outlines indicating greater weight. Dashed regions and crosses show
		Louvain communities and their centroids.}
		\label{fig:rules}
	\end{figure}

	We modify the original embeddings offline to incorporate information from related nodes
	in the graph. The five transformations differ in which nodes influence each modified
	vector and how their contributions are combined (Figure~\ref{fig:rules}).

	Original embeddings encode chunk text or, for document and entity nodes, their names.
	We embed nodes and queries with OpenAI's \texttt{text-embedding-3-small}, using $1{,}536$ dimensions.
	Each transformation produces one modified vector
	per node. We denote the original and modified embeddings of node $i$ by $\mathbf{x}_i$
	and $\mathbf{z}_i$, respectively. Appendix~\ref{app:retro} lists all evaluated transformations
	and hyperparameter configurations.

	\paragraph{One-hop neighbour blend (hop-1).}
    For each node $i$, we consider its one-hop neighbours. We define $S_i$ as the $20\,\%$ of these nodes with the lowest cosine similarity to the original embedding $\mathbf{x}_i$, selecting at least two whenever available. The modified embedding $\mathbf{z}_i$ combines the original vector with the mean of the original embeddings of the selected neighbours:
	\begin{equation}
		\mathbf{z}_i =
		\alpha \mathbf{x}_i +
		(1-\alpha)\frac{1}{|S_i|}
		\sum_{j\in S_i}\mathbf{x}_j,
	\end{equation}
	where $\alpha=0.2$. Selecting the least semantically similar one-hop neighbours is intended to introduce graph-related information that is not already encoded in the source representation.

	\paragraph{Two-hop neighbour blend (hop-2).}
	This method uses the same selection and blending procedure, applied to nodes at graph distance exactly $2$.

	\paragraph{Convex neighbour pull.}
    Let $\mathbf{Z}$ collect the modified node embeddings and $E$ denote the graph's edge set.
    We minimise the classical retrofitting objective~\cite{faruqui} with uniform anchor and edge weights:
\begin{equation}
	J(\mathbf{Z}) =
	\sum_i \lVert \mathbf{z}_i-\mathbf{x}_i\rVert_2^2
	+
	\sum_{(i,j)\in E}
	\lVert \mathbf{z}_i-\mathbf{z}_j\rVert_2^2 .
\end{equation}
The first term anchors each node to its original embedding, while the second pulls adjacent nodes closer together in the modified space. The anchoring term makes the objective strictly convex, ensuring a unique global minimiser to which the iterative optimisation procedure converges.

	\paragraph{Random-walk endpoint blend (walk).}
	This transformation launches 100 random walks of exactly two steps from each node. The modified embedding is computed as the frequency-weighted mean of the endpoints' original embeddings. 

	\paragraph{Louvain cluster blend.}
	This transformation partitions the unweighted, undirected graph using Louvain community detection at resolution $0.5$; we refer to Blondel et al.~\cite{blondel} for details of the community-construction algorithm. Each node $i$ is then moved towards the centroid $\boldsymbol{\mu}_{c(i)}$ of its community $c(i)$:
	\begin{equation}
		\mathbf{z}_i =
		\alpha\mathbf{x}_i +
		(1-\alpha)\boldsymbol{\mu}_{c(i)} .
	\end{equation}
	Each centroid is the mean of its community's original embeddings and remains fixed. The selected configuration uses $\alpha=0.5$, assigning equal weight to the node's original embedding and its community centroid.


	\subsection{Retrieval Strategies}\label{retr-sect}

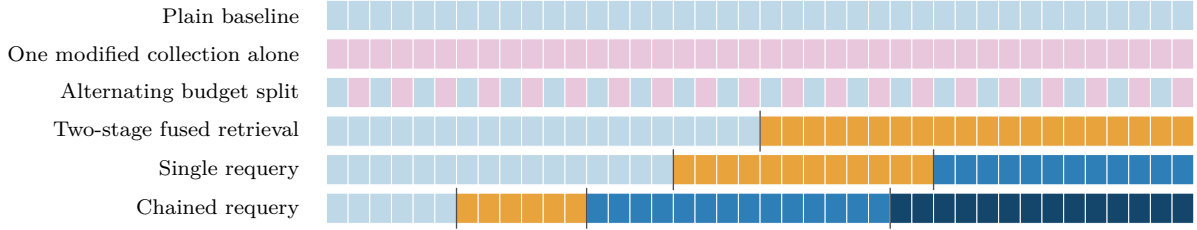
\begin{figure}[H]
		\centering
		\begin{tikzpicture}[x=1.85cm, y=2cm, font=\scriptsize,
			cell/.style={draw=white, line width=0.25pt},
			rule/.style={black!70, line width=0.5pt},
			rlab/.style={anchor=east, font=\scriptsize},
			hd/.style={font=\scriptsize, text=black!65}]
			\node[rlab] at (-0.12,-0.100) {Plain baseline};
			\foreach \i in {0,...,39}{\fill[colorig,cell] (\i*0.155,-0.000) rectangle (\i*0.155+0.155,-0.200);}
			\node[rlab] at (-0.12,-0.355) {One modified collection alone};
			\foreach \i in {0,...,39}{\fill[colmod,cell] (\i*0.155,-0.255) rectangle (\i*0.155+0.155,-0.455);}
			\node[rlab] at (-0.12,-0.610) {Alternating budget split};
			\foreach \i in {0,2,...,38}{\fill[colorig,cell] (\i*0.155,-0.510) rectangle (\i*0.155+0.155,-0.710);}
			\foreach \i in {1,3,...,39}{\fill[colmod,cell] (\i*0.155,-0.510) rectangle (\i*0.155+0.155,-0.710);}
			\node[rlab] at (-0.12,-0.865) {Two-stage fused retrieval};
			\foreach \i in {0,...,19}{\fill[colorig,cell] (\i*0.155,-0.765) rectangle (\i*0.155+0.155,-0.965);}
			\foreach \i in {20,...,39}{\fill[colfused,cell] (\i*0.155,-0.765) rectangle (\i*0.155+0.155,-0.965);}
			\draw[rule] (3.100,-0.730) -- (3.100,-1.000);
			\node[rlab] at (-0.12,-1.120) {Single requery};
			\foreach \i in {0,...,15}{\fill[colorig,cell] (\i*0.155,-1.020) rectangle (\i*0.155+0.155,-1.220);}
			\foreach \i in {16,...,27}{\fill[colfused,cell] (\i*0.155,-1.020) rectangle (\i*0.155+0.155,-1.220);}
			\foreach \i in {28,...,39}{\fill[colshift,cell] (\i*0.155,-1.020) rectangle (\i*0.155+0.155,-1.220);}
			\draw[rule] (2.480,-0.985) -- (2.480,-1.255);
			\draw[rule] (4.340,-0.985) -- (4.340,-1.255);
			\node[rlab] at (-0.12,-1.375) {Chained requery};
			\foreach \i in {0,...,5}{\fill[colorig,cell] (\i*0.155,-1.275) rectangle (\i*0.155+0.155,-1.475);}
			\foreach \i in {6,...,11}{\fill[colfused,cell] (\i*0.155,-1.275) rectangle (\i*0.155+0.155,-1.475);}
			\foreach \i in {12,...,25}{\fill[colshift,cell] (\i*0.155,-1.275) rectangle (\i*0.155+0.155,-1.475);}
			\foreach \i in {26,...,39}{\fill[colshifttwo,cell] (\i*0.155,-1.275) rectangle (\i*0.155+0.155,-1.475);}
			\draw[rule] (0.930,-1.240) -- (0.930,-1.510);
			\draw[rule] (1.860,-1.240) -- (1.860,-1.510);
			\draw[rule] (4.030,-1.240) -- (4.030,-1.510);
		\end{tikzpicture}
		\caption{Allocation of the 40-node retrieval budget. Each cell represents one returned
		node. Light blue and pink denote original- and modified-space searches; orange denotes
		fused scoring. The two darker blues denote successive searches with shifted queries
		in the original space. Vertical lines separate stages.}
		\label{fig:strategies}
	\end{figure}

	We compare retrieval strategies that use the original and graph-modified embeddings
	separately or together. These range from direct searches in either space to combinations
	of independent rankings and multi-stage searches guided by earlier results. We hold the
	output at $40$ distinct nodes to compare evidence recovery under the same retrieval budget
	(Figure~\ref{fig:strategies}).

	All strategies use cosine similarity over stored embeddings of all node types, without
	graph access at query time. Later stages exclude previously returned nodes.
	Appendix~\ref{app:retr} lists all evaluated methods and configurations.

\paragraph{Direct retrieval.}
The first two rows of Figure~\ref{fig:strategies} each use a single embedding collection.
The \emph{plain baseline} returns the nodes whose original embeddings are most similar
to the query embedding $\mathbf{q}$. \emph{One modified collection alone} applies the same search
to graph-modified embeddings. This comparison tests whether incorporating graph structure
into the vectors makes them more useful for direct query matching.

\paragraph{Alternating budget split.}
This strategy gives both spaces a share of the output. We search them independently
using $\mathbf{q}$ and alternate between their rankings, skipping duplicate node identifiers.

\paragraph{Two-stage fused retrieval.}
Here, the initial results guide the search for further evidence. The first stage retrieves
20 nodes (seeds) from the original space using $\mathbf{q}$ and computes
$\mathbf{m}$ as the mean of their modified embeddings. The second stage pools candidates
from an original-space search with $\mathbf{q}$ and a modified-space search with $\mathbf{m}$,
then ranks each candidate node $n$ by the \emph{fused score}
\begin{equation}\label{eq:fused-score}
	s(n,\alpha)
	=
	(1-\alpha)\cos(\mathbf{q},\mathbf{x}_n)
	+
	\alpha\cos(\mathbf{m},\mathbf{z}_n).
\end{equation}
The fusion weight $\alpha=0.5$ gives equal weight to similarity to the original query and
similarity to the seeds in the modified space. The 20 highest-scoring remaining nodes
complete the output.

\paragraph{Single requery.}
This strategy uses the fused-stage results to refine the query for another search in the
original space. It retrieves 16 seeds, selects 12 nodes using the fused score in Equation~\eqref{eq:fused-score}, and averages
the latter's original embeddings to obtain $\bar{\mathbf{x}}$. The shifted query
\begin{equation}\label{eq:query-update}
	\mathbf{q}'=0.7\mathbf{q}+0.3\bar{\mathbf{x}}
\end{equation}
retrieves the final 12 nodes, giving a $16/12/12$ allocation.

\paragraph{Chained requery.}
This strategy adds a second query update, so the first shifted search supplies evidence
for the next. It uses $6/6/14/14$ nodes for seed retrieval, fused scoring, and two
shifted-query searches in the original space. Each update uses the mean original embedding
of the preceding stage's results. Both use Equation~\eqref{eq:query-update} with the original query $\mathbf{q}$; only the
mean $\bar{\mathbf{x}}$ changes.

	\paragraph{Controls without graph conditioning.}
	To separate the contribution of graph conditioning from that of query shifting, we compare
	each requery strategy with two controls that change only its second stage. One sets
	the fusion weight in Equation~\eqref{eq:fused-score} to $\alpha=0$; the other retains $\alpha=0.5$ but uses original embeddings for both the seed
	mean $\mathbf{m}$ and the candidates. Section~\ref{sec:graph-conditioning} presents this comparison.

%% file: src/experiments.tex
	\section{Experimental Results}\label{sec:experiments}
	We first test graph-modified embeddings in direct retrieval, then compare strategies
	that combine original and modified representations. Finally, control experiments assess
	the contribution of graph conditioning.

	For each question, gold-evidence recovery is the fraction of its gold chunks returned;
	reported percentages average this fraction equally across questions. Document and entity
	nodes consume retrieval budget but do not count as recovered evidence.
	\subsection{Direct Retrieval with Graph-Modified Embeddings}
	We compare the original embeddings with the five graph-modified collections as direct
	search representations across all $977$ validation questions.
	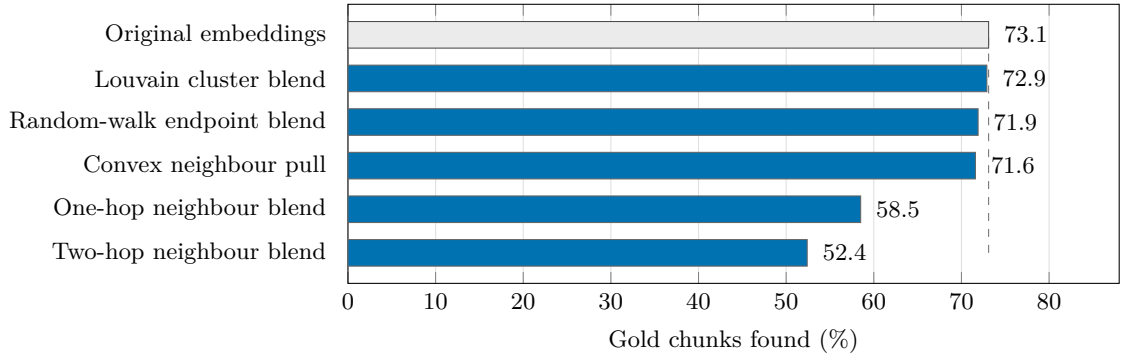
\begin{figure}[H]
		\centering
		\begin{tikzpicture}
			\begin{axis}[
				xbar,
				width=0.74\linewidth, height=150pt,
				xmin=0, xmax=88,
				xlabel={Gold chunks found (\%)},
				symbolic y coords={Two-hop neighbour blend,One-hop neighbour blend,Convex neighbour pull,Random-walk endpoint blend,
					Louvain cluster blend,Original embeddings},
				ytick={Two-hop neighbour blend,One-hop neighbour blend,Convex neighbour pull,Random-walk endpoint blend,Louvain cluster blend,%
					Original embeddings},
				bar width=10pt,
				bar shift=0pt,
				enlarge y limits=0.14,
				nodes near coords={\pgfmathprintnumber[fixed,precision=1]{\pgfplotspointmeta}},
				nodes near coords style={font=\footnotesize, anchor=west, xshift=2pt},
				xmajorgrids, grid style={black!12},
				tick label style={font=\footnotesize},
				label style={font=\footnotesize},
				ytick style={draw=none},
				]
				\draw[dashed, black!55]
				(axis cs:73.1,Two-hop neighbour blend) -- (axis cs:73.1,Original embeddings);
				\addplot[draw=black!60, fill=colwin, bar shift=0pt] coordinates {
					(52.4,Two-hop neighbour blend) (58.5,One-hop neighbour blend) (71.6,Convex neighbour pull)
					(71.9,Random-walk endpoint blend) (72.9,Louvain cluster blend)};
				\addplot[draw=black!70, fill=black!8, bar shift=0pt] coordinates {
					(73.1,Original embeddings)};
			\end{axis}
		\end{tikzpicture}
		\caption{Gold-evidence recovery under direct retrieval across all $977$ validation questions. The dashed line denotes plain vector retrieval.}
		\label{fig:chunks-found}
	\end{figure}
	All five transformations recover fewer gold chunks than the original embeddings,
	with the largest losses observed for the one-hop and two-hop neighbour blends.
\subsection{Comparison of Retrieval Strategies}
\begin{figure}[H]
	\centering
	\begin{tikzpicture}
		\begin{axis}[
			width=0.78\linewidth, height=138pt,
			enlarge x limits=0.10,
			ymin=20, ymax=65, ytick={20,30,40,50,60},
			yticklabel={\pgfmathprintnumber{\tick}\,\%},
			ylabel={Gold chunks found (\%)},
			symbolic x coords={alone,split,fused,single,chained}, xtick=data,
			xticklabels={{One\\alone},{Split},{Fused},{Single\\requery},{Chained\\requery}},
			x tick label style={font=\scriptsize, align=center},
			tick label style={font=\footnotesize}, label style={font=\footnotesize},
			ymajorgrids, grid style={black!12},
			legend style={font=\footnotesize, at={(0.5,1.02)}, anchor=south,
				legend columns=6, draw=none, column sep=5pt},
			]
			\addplot[colconvex,  thick, mark=*] coordinates {(alone,47.5) (split,48.6) (fused,54.0) (single,56.7) (chained,58.2)};
			\addplot[collouvain, thick, mark=square*] coordinates {(alone,51.5) (split,50.7) (fused,55.4) (single,56.3) (chained,55.9)};
			\addplot[colwalk,    thick, mark=triangle*] coordinates {(alone,49.8) (split,52.3) (fused,56.0) (single,58.6) (chained,60.8)};
			\addplot[colhoptwo,  thick, mark=diamond*] coordinates {(alone,39.0) (split,55.1) (fused,55.6) (single,58.1) (chained,59.2)};
			\addplot[colhopone,  thick, mark=pentagon*] coordinates {(alone,26.6) (split,41.4) (fused,54.5) (single,57.5) (chained,59.1)};
			\draw[dashed, black!65] ({rel axis cs:0,0}|-{axis cs:alone,51.0})
			-- ({rel axis cs:1,0}|-{axis cs:alone,51.0});
			\addlegendimage{dashed, black!65}
			\legend{convex, Louvain, walk, hop-2, hop-1, Plain baseline}
		\end{axis}
	\end{tikzpicture}
	\caption{Gold-evidence recovery across retrieval strategies on the $527$ difficult questions. The dashed line denotes plain retrieval.}
	\label{fig:hard-stages}
\end{figure}
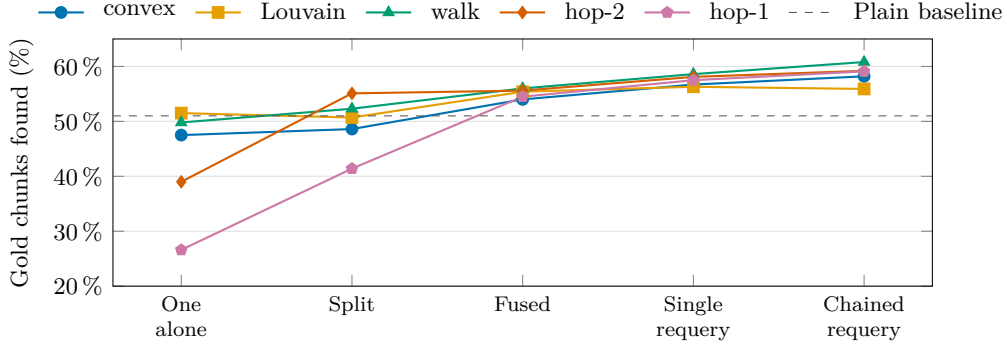

Plain retrieval already recovers $98.9\,\%$ of the gold evidence on the $450$
single-fact questions, leaving little room to distinguish strategies. Unless otherwise
stated, the remaining comparisons therefore focus on the $527$ difficult questions in the other two groups.
Figure~\ref{fig:hard-stages} compares the five embedding
transformations across retrieval strategies.

Performance varies most across embedding transformations under direct retrieval and
budget splitting. Fused retrieval and requery outperform the plain baseline with all five
transformations, with smaller differences between them.

%% file: src/retrieval_analysis.tex
	\subsection{Alternating Budget Split}
	The alternating split assigns half of the retrieval budget to each embedding space.
	In the group-level plots, error bars show $95\,\%$ confidence intervals for changes
	relative to the baseline, estimated from $10{,}000$ paired bootstrap resamples of questions.
	\begin{figure}[H]
		\providecolor{colhoptwo}{HTML}{CC79A7}
		\centering
		\begin{subfigure}[t]{0.51\textwidth}
			\centering
			\begin{tikzpicture}
				\begin{axis}[
					xbar,
					scale only axis,
					width=0.60\linewidth, height=118pt,
					xmin=0, xmax=100,
					xtick={0,25,50,75,100},
					xlabel={Gold chunks found (\%)},
					symbolic y coords={hop1,convex,louvain,originals,walk,hop2},
					ytick={hop1,convex,louvain,originals,walk,hop2},
					yticklabels={hop-1,convex,Louvain,plain baseline,walk,hop-2},
					bar width=7pt,
					bar shift=0pt,
					enlarge y limits=0.12,
					nodes near coords={\pgfmathprintnumber[fixed,fixed zerofill,precision=1]{\pgfplotspointmeta}},
					nodes near coords style={font=\scriptsize, anchor=west, xshift=2pt},
					xmajorgrids, grid style={black!12},
					tick label style={font=\scriptsize},
					label style={font=\footnotesize},
					ytick style={draw=none},
					]
					\draw[dashed, black!55] (axis cs:51.0,hop1) -- (axis cs:51.0,hop2);
					\addplot[draw=black!60, fill=colwin, bar shift=0pt] coordinates {
						(41.4,hop1) (48.6,convex) (50.7,louvain) (52.3,walk) (55.1,hop2)};
					\addplot[draw=black!70, fill=black!8, bar shift=0pt] coordinates {
						(51.0,originals)};
				\end{axis}
			\end{tikzpicture}
			\caption{Gold-evidence recovery under the alternating budget split for each graph-modified embedding collection. The dashed line denotes the performance of plain vector retrieval.}
			\label{fig:chunks-found-split}
		\end{subfigure}
		\hfill
		\begin{subfigure}[t]{0.47\textwidth}
	\centering
	\begin{tikzpicture}
		\begin{axis}[
			xbar,
			scale only axis,
			width=0.48\linewidth, height=112pt,
			bar width=9pt,
			bar shift=0pt,
			xmin=-16, xmax=30,
			clip=false,
			xtick={-10,0,10,20},
			xlabel={$\Delta$ vs.\ the baseline (points)},
			symbolic y coords={extremize,multihop},
			ytick={extremize,multihop},
			yticklabels={{Extremisation\\[-1pt]{\scriptsize 59.6\,\%}},{Relational and\\collection\\[-1pt]{\scriptsize 42.9\,\%}}},
			y tick label style={font=\scriptsize, align=right},
			enlarge y limits=0.28,
			tick label style={font=\scriptsize}, label style={font=\footnotesize},
			xmajorgrids, grid style={black!12},
			ytick style={draw=none},
			]
			\draw[black!55, line width=0.7pt] (axis cs:0,extremize) -- (axis cs:0,multihop);
			\addplot[draw=black!60, fill=colwin, bar shift=0pt,
			error bars/.cd, x dir=both, x explicit,
			error bar style={black!55, line width=0.5pt}, error mark options={rotate=90, black!55, mark size=1.1pt}]
			coordinates {
				(-10.4,extremize) -= (2.9,0) += (3.0,0)
				(17.9,multihop) -= (2.7,0) += (2.6,0)
			};
			\node[font=\scriptsize, anchor=west, xshift=6pt, align=left] at ({rel axis cs:1,0} |- {axis cs:0,extremize}) {$-10.4$\\[-2pt]{\scriptsize\textcolor{black!55}{-17.4\,\%}}};
			\node[font=\scriptsize, anchor=west, xshift=6pt, align=left] at ({rel axis cs:1,0} |- {axis cs:0,multihop}) {$+17.9$\\[-2pt]{\scriptsize\textcolor{black!55}{+41.6\,\%}}};
		\end{axis}
	\end{tikzpicture}
    \caption{Effect of the hop-2 alternating budget split across question groups. Baseline recovery is shown below each group label; bar annotations report the corresponding absolute and relative changes.}
	\label{fig:split-by-group}
\end{subfigure}

		\caption{Alternating budget split on the $527$ difficult questions.}
		\label{fig:split-pair}
	\end{figure}
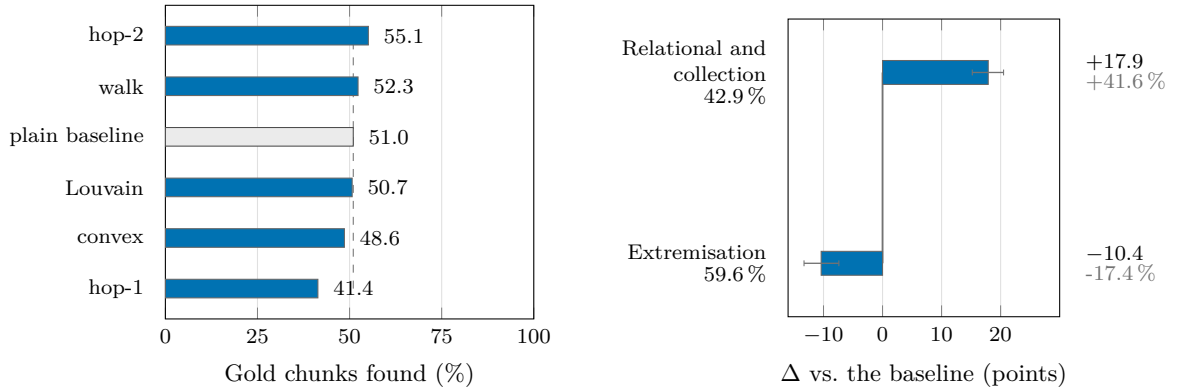
The alternating budget split reveals a clear trade-off. Panel~(a) shows that only the walk and hop-2 representations outperform plain retrieval, with hop-2 performing best. However, Panel~(b) shows that its gain is uneven: relational and collection questions improve by $41.6\,\%$, whereas extremisation questions decline by $17.4\,\%$ relative to their baselines. The modified space helps reach semantically distant but graph-connected evidence, while assigning half of the budget to it reduces the broad candidate coverage required for extremisation.
	\subsection{Two-Stage Fused Retrieval}
	We next test whether letting the original-space results guide the second stage yields
	more consistent gains than the independent budget split.
	\begin{figure}[H]
		\providecolor{colhoptwo}{HTML}{CC79A7}\providecolor{colwin}{HTML}{0072B2}
		\centering
		\begin{subfigure}[t]{0.55\textwidth}
			\centering
			\begin{tikzpicture}
				\begin{axis}[
					xbar,
					scale only axis,
					width=0.64\linewidth, height=125pt,
					xmin=0, xmax=100,
					xtick={0,25,50,75,100},
					xlabel={Gold chunks found (\%)},
					symbolic y coords={originals,convex,hop1,louvain,hop2,walk},
					ytick={originals,convex,hop1,louvain,hop2,walk},
					yticklabels={plain baseline,convex,hop-1,Louvain,hop-2,walk},
					bar width=7pt,
					bar shift=0pt,
					enlarge y limits=0.12,
					nodes near coords={\pgfmathprintnumber[fixed,fixed zerofill,precision=1]{\pgfplotspointmeta}},
					nodes near coords style={font=\scriptsize, anchor=west, xshift=2pt},
					xmajorgrids, grid style={black!12},
					tick label style={font=\scriptsize},
					label style={font=\footnotesize},
					ytick style={draw=none},
					]
					\draw[dashed, black!55] (axis cs:51.0,originals) -- (axis cs:51.0,walk);
					\addplot[draw=black!60, fill=colwin, bar shift=0pt] coordinates {
						(54.0,convex) (54.5,hop1) (55.4,louvain) (55.6,hop2) (56.0,walk)};
					\addplot[draw=black!70, fill=black!8, bar shift=0pt] coordinates {
						(51.0,originals)};
				\end{axis}
			\end{tikzpicture}
			\caption{Gold-evidence recovery of the two-stage fused retriever across graph-modified embedding collections. The dashed line denotes plain vector retrieval.}
			\label{fig:chunks-found-twostage}
		\end{subfigure}
		\hfill
			\begin{subfigure}[t]{0.43\textwidth}
		\centering
		\begin{tikzpicture}
			\begin{axis}[
				xbar,
				scale only axis,
				width=0.43\linewidth, height=112pt,
				bar width=9pt,
				bar shift=0pt,
				xmin=-6, xmax=14,
				clip=false,
				xtick={-5,0,5,10},
				xlabel={$\Delta$ vs.\ the baseline (points)},
				symbolic y coords={extremize,multihop},
				ytick={extremize,multihop},
				yticklabels={{Extremisation\\[-1pt]{\scriptsize 59.6\,\%}},{Relational and\\collection\\[-1pt]{\scriptsize 42.9\,\%}}},
				y tick label style={font=\scriptsize, align=right},
				enlarge y limits=0.28,
				tick label style={font=\scriptsize}, label style={font=\footnotesize},
				xmajorgrids, grid style={black!12},
				ytick style={draw=none},
				]
				\draw[black!55, line width=0.7pt] (axis cs:0,extremize) -- (axis cs:0,multihop);
				\addplot[draw=black!60, fill=colwin, bar shift=0pt,
				error bars/.cd, x dir=both, x explicit,
				error bar style={black!55, line width=0.5pt}, error mark options={rotate=90, black!55, mark size=1.1pt}]
				coordinates {
					(1.3,extremize) -= (3.2,0) += (3.2,0)
					(7.6,multihop) -= (1.8,0) += (1.8,0)
				};
				\node[font=\scriptsize, anchor=west, xshift=6pt, align=left] at ({rel axis cs:1,0} |- {axis cs:0,extremize}) {$+1.3$\\[-2pt]{\scriptsize\textcolor{black!55}{+2.2\,\%}}};
				\node[font=\scriptsize, anchor=west, xshift=6pt, align=left] at ({rel axis cs:1,0} |- {axis cs:0,multihop}) {$+7.6$\\[-2pt]{\scriptsize\textcolor{black!55}{+17.6\,\%}}};
			\end{axis}
		\end{tikzpicture}
		\caption{Effect of two-stage fused retrieval with the hop-2 representation across question groups. Baseline recovery is shown below each group label; bar annotations report absolute and relative improvements.}
		\label{fig:fused-by-group}
	\end{subfigure}

		\caption{Two-stage fused retrieval on the $527$ difficult questions.}
		\label{fig:twostage-pair}
	\end{figure}
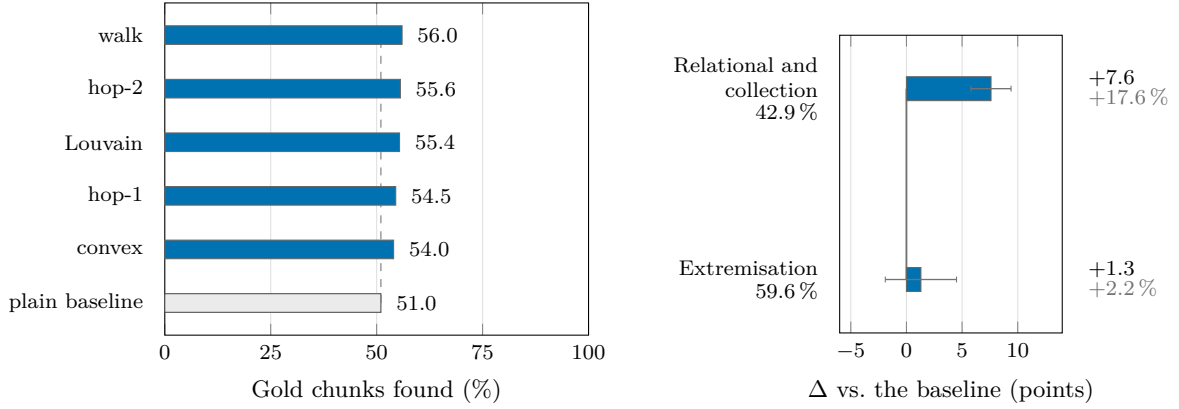

	\needspace{3\baselineskip}
	Two-stage fused retrieval improves over the baseline with every graph-modified collection, recovering $54.0\,\%$ to $56.0\,\%$ of the gold evidence against $51.0\,\%$ for plain retrieval. The narrow range suggests less sensitivity to the choice of transformation than under budget splitting. The walk representation performs best, gaining $5.0$ percentage points ($9.8\,\%$ relative).

	\needspace{3\baselineskip}
	Panel~(b) again uses hop-2 to compare directly with the alternating split. Its gain is concentrated on relational and collection questions: $7.6$ points ($17.6\,\%$ relative). The improvement on extremisation questions is small and statistically uncertain.
	\subsection{Requery Strategies and Graph Conditioning}\label{sec:requery-analysis}
\subsubsection{Performance of Requery Strategies}
	We examine whether using the fused-stage results to refine the query improves recovery
	beyond the two-stage schedule.
\begin{figure}[H]
	\centering
	\begin{tikzpicture}
		\begin{axis}[
			xbar,
			scale only axis,
			width=0.60\linewidth, height=168pt,
			xmin=0, xmax=100,
			xtick={0,25,50,75,100},
			xlabel={Gold chunks found (\%)},
			symbolic y coords={louvain,convex,hop1,hop2,walk,originals},
			ytick={louvain,convex,hop1,hop2,walk,originals},
			yticklabels={Louvain,convex,hop-1,hop-2,walk,plain baseline},
			bar width=7pt,
			enlarge y limits=0.14,
			nodes near coords={%
				\pgfmathfloattofixed{\pgfplotspointmeta}\let\barval\pgfmathresult
				\pgfmathsetmacro{\bardelta}{100*(\barval-51.0)/51.0}%
				\pgfmathprintnumber[fixed,fixed zerofill,precision=1]{\barval}%
				\ifdim\bardelta pt>0.05pt\relax
				\,\textcolor{black!55}{($+$\pgfmathprintnumber[fixed,fixed zerofill,precision=1]{\bardelta}\,\%)}%
				\fi},
			nodes near coords style={font=\scriptsize, anchor=west, xshift=2pt},
			xmajorgrids, grid style={black!12},
			tick label style={font=\scriptsize},
			label style={font=\footnotesize},
			legend style={font=\scriptsize, at={(0.5,1.03)}, anchor=south, legend columns=3, draw=none},
			legend image code/.code={\draw[#1] (0cm,-0.08cm) rectangle (0.22cm,0.14cm);},
			ytick style={draw=none},
			]
			\draw[dashed, black!55] (axis cs:51.0,louvain) -- (axis cs:51.0,originals);
			\addplot[draw=black!60, fill=colwin, bar shift=-5.5pt] coordinates {
				(56.3,louvain) (56.7,convex) (57.5,hop1) (58.1,hop2) (58.6,walk)};
			\addplot[draw=black!60, fill=colshifttwo, bar shift=5.5pt] coordinates {
				(55.9,louvain) (58.2,convex) (59.1,hop1) (59.2,hop2) (60.8,walk)};
			\addplot[draw=black!70, fill=black!8, bar shift=0pt] coordinates {
				(51.0,originals)};
			\legend{single requery, chained requery, plain baseline}
		\end{axis}
	\end{tikzpicture}
	\caption{Gold-evidence recovery of the single requery and chained requery schedules across graph-modified embedding collections. The dashed line denotes plain vector retrieval.}
	\label{fig:requery-vs-baseline}
\end{figure}
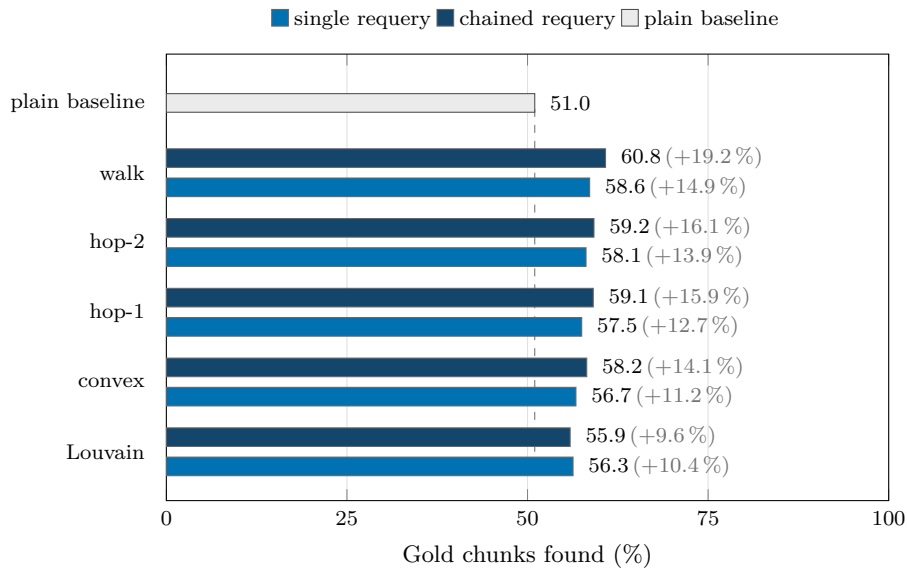
Chained requery with the walk representation performs best, recovering $60.8\,\%$ of the gold evidence against $51.0\,\%$ for plain retrieval, a gain of $9.8$ percentage points ($19.2\,\%$ relative). The walk representation also performs best with single requery.

\begin{figure}[H]
	\providecolor{colshifttwo}{HTML}{14456B}
	\centering
	\begin{tikzpicture}
		\begin{axis}[
			xbar,
			scale only axis,
			width=0.52\linewidth, height=95pt,
			bar width=9pt,
			bar shift=0pt,
			xmin=-2, xmax=29,
			clip=false,
			xtick={0,5,10,15,20},
			xlabel={$\Delta$ gold-evidence recovery vs.\ the plain baseline (points)},
			symbolic y coords={extremize,multihop},
			ytick={extremize,multihop},
			yticklabels={{Extremisation\\questions\\[-1pt]{\scriptsize baseline 59.6\,\%}},{Relational and\\collection\\[-1pt]{\scriptsize baseline 42.9\,\%}}},
			y tick label style={font=\footnotesize, align=right},
			enlarge y limits=0.28,
			tick label style={font=\footnotesize}, label style={font=\footnotesize},
			xmajorgrids, grid style={black!12},
			ytick style={draw=none},
			]
			\draw[black!55, line width=0.7pt] (axis cs:0,extremize) -- (axis cs:0,multihop);
			\addplot[draw=black!60, fill=colshifttwo, bar shift=0pt,
			error bars/.cd, x dir=both, x explicit,
			error bar style={black!55, line width=0.5pt}, error mark options={rotate=90, black!55, mark size=1.2pt}]
			coordinates {
				(15.4,extremize) -= (2.5,0) += (2.7,0)
				(4.3,multihop) -= (1.6,0) += (1.6,0)
			};
			\node[font=\scriptsize, anchor=west] at (axis cs:18.7,extremize) {$+15.4$\ \textcolor{black!55}{(+25.9\,\%)}};
			\node[font=\scriptsize, anchor=west] at (axis cs:6.5,multihop) {$+4.3$\ \textcolor{black!55}{(+10.0\,\%)}};
		\end{axis}
	\end{tikzpicture}
	\caption{Improvement of chained requery with the walk representation over plain retrieval, reported by question group. Annotations show absolute and relative gains.}
	\label{fig:chained-by-group}
\end{figure}
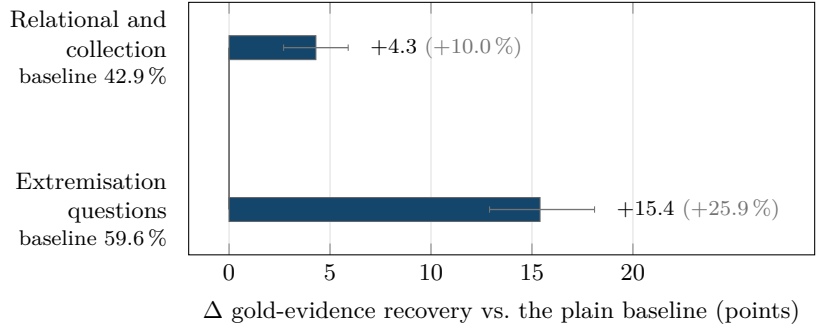

Unlike the alternating split, this configuration improves both difficult question groups (Figure~\ref{fig:chained-by-group}). Recovery increases by $15.4$ points on extremisation questions ($25.9\,\%$ relative) and by $4.3$ points on relational and collection questions ($10.0\,\%$ relative).





\subsubsection{Contribution of Graph Conditioning}\label{sec:graph-conditioning}
The fused stage selects the nodes used for subsequent query shifts, so graph conditioning can affect later searches as well as the second-stage results.

We compare chained requery with two controls using the same $6/6/14/14$ schedule. The first sets the fusion weight in Equation~\eqref{eq:fused-score} to $\alpha=0$, reducing the second-stage score to $\cos(\mathbf q,\mathbf x_n)$. The second uses original embeddings for both seeds and candidates:
\begin{equation}
	0.5\cdot\cos(\mathbf q,\mathbf x_n)
	+0.5\cdot\cos(\mathbf m,\mathbf x_n),
\end{equation}
where $\mathbf m$ is the mean original embedding of the first-stage results. These controls measure the benefit of the multi-stage schedule and query reformulation without graph conditioning.
\begin{figure}[H]
	\providecolor{colwin}{HTML}{0072B2}\providecolor{colctrl}{HTML}{E69F00}\providecolor{colrot}{HTML}{C0812A}
	\centering
	\begin{tikzpicture}
		\begin{axis}[
			scale only axis, width=0.62\linewidth, height=128pt,
			xlabel={Nodes returned}, ylabel={Gold chunks found (\%)},
			xmin=5.2, xmax=40.8, ymin=18, ymax=64,
			xtick={6,12,20,26,33,40}, ytick={20,30,40,50,60},
			tick label style={font=\scriptsize}, label style={font=\scriptsize},
			grid=major, grid style={black!10},
			legend columns=2, legend cell align=left,
			legend style={font=\scriptsize, draw=none, at={(0.5,-0.22)}, anchor=north,
				/tikz/every even column/.append style={column sep=7pt}},
			]
			\draw[black!35, densely dotted] (axis cs:12,18) -- (axis cs:12,64);
			\draw[black!35, densely dotted] (axis cs:26,18) -- (axis cs:26,64);
			\node[font=\tiny, black!45, anchor=south] at (axis cs:9,61.2) {2};
			\node[font=\tiny, black!45, anchor=south] at (axis cs:19,61.2) {3};
			\node[font=\tiny, black!45, anchor=south] at (axis cs:33,61.2) {4};
			\draw[black!40, line width=0.4pt, loosely dashed] (axis cs:5.2,51.03) -- (axis cs:40.8,51.03);
			\addplot[black!65, line width=0.9pt, dashed, mark=none] coordinates {
				(6,21.38) (7,23.33) (8,25.08) (9,26.66) (10,28.23) (11,30.07) (12,31.42)
				(13,32.43) (14,33.61) (15,34.54) (16,35.24) (17,36.35) (18,37.34) (19,38.30)
				(20,38.83) (21,39.62) (22,40.53) (23,41.00) (24,42.02) (25,42.67) (26,43.33)
				(27,43.70) (28,44.46) (29,45.05) (30,45.47) (31,45.96) (32,46.58) (33,47.03)
				(34,47.53) (35,48.12) (36,49.33) (37,49.70) (38,50.31) (39,50.64) (40,51.03)
			};
			\addplot[colctrl, line width=1.1pt, mark=none] coordinates {
				(6,21.38) (7,23.33) (8,25.08) (9,26.66) (10,28.23) (11,30.07) (12,31.42)
				(13,32.63) (14,33.69) (15,34.78) (16,36.16) (17,37.00) (18,38.00) (19,38.88)
				(20,39.76) (21,40.78) (22,41.33) (23,42.16) (24,43.24) (25,44.05) (26,45.01)
				(27,45.86) (28,46.67) (29,47.70) (30,48.28) (31,48.83) (32,49.43) (33,50.19)
				(34,50.52) (35,51.13) (36,51.76) (37,52.60) (38,53.22) (39,53.83) (40,54.71)
			};
			\addplot[colrot, line width=1.5pt, dash pattern=on 0.1pt off 3pt, line cap=round, mark=none] coordinates {
				(6,21.38) (7,22.98) (8,25.46) (9,27.19) (10,29.07) (11,30.73) (12,32.77)
				(13,34.12) (14,35.84) (15,37.09) (16,38.18) (17,39.02) (18,40.07) (19,40.96)
				(20,41.73) (21,42.41) (22,43.19) (23,43.88) (24,44.53) (25,45.37) (26,45.99)
				(27,46.92) (28,47.55) (29,48.32) (30,49.05) (31,49.73) (32,50.44) (33,51.33)
				(34,52.03) (35,52.73) (36,53.12) (37,53.82) (38,54.53) (39,54.99) (40,55.46)
			};
			\addplot[colwin, line width=1.3pt, mark=none] coordinates {
				(6,21.38) (7,25.40) (8,29.58) (9,32.83) (10,35.58) (11,37.82) (12,39.58)
				(13,40.88) (14,42.60) (15,43.90) (16,44.66) (17,45.67) (18,46.42) (19,47.22)
				(20,48.13) (21,48.91) (22,49.69) (23,50.53) (24,51.52) (25,52.58) (26,53.22)
				(27,53.81) (28,54.45) (29,54.93) (30,55.63) (31,56.16) (32,56.61) (33,57.10)
				(34,57.99) (35,58.45) (36,58.94) (37,59.54) (38,59.99) (39,60.44) (40,60.75)
			};
			\legend{{plain baseline}, {control ($\alpha=0$)},
				{control (original space)}, {chained requery (walk)}}
		\end{axis}
	\end{tikzpicture}
	\caption{Cumulative gold-evidence recovery over the first $k$ nodes returned in stage order, without global reranking. The full chained requery method uses the walk representation; both controls retain its $6/6/14/14$ schedule but remove graph conditioning. Vertical dotted lines mark stage boundaries. The horizontal dashed line shows plain retrieval's recovery at 40 nodes.}
	\label{fig:depth-pair}
\end{figure}
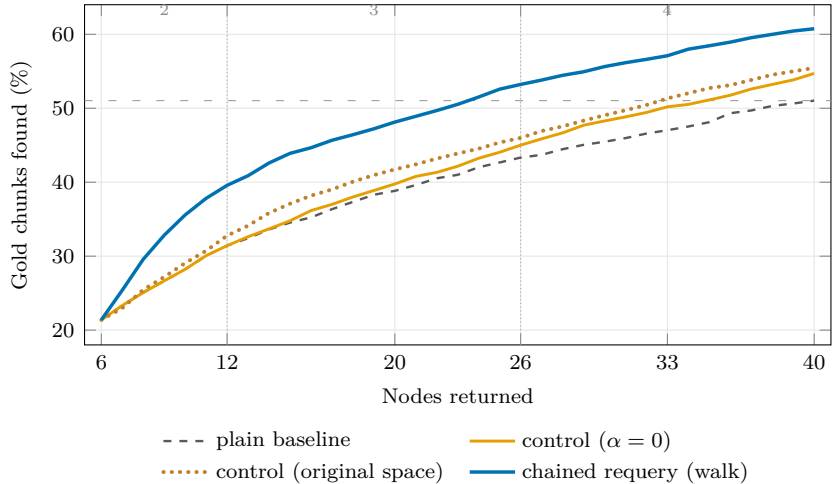
\needspace{3\baselineskip}
At the full budget, the controls recover $54.7\,\%$ and $55.5\,\%$ of the gold evidence, compared with $51.0\,\%$ for plain retrieval (Figure~\ref{fig:depth-pair}). Original-space fusion provides only a small improvement over query-only scoring in this configuration.

The full method reaches $60.8\,\%$, exceeding the stronger control by $5.3$ points. Graph conditioning thus contributes slightly more than half of its $9.8$-point gain over plain retrieval. The advantage emerges during the fused stage and persists through later searches. After twelve returned nodes, recovery is $39.6\,\%$, against $31.4\,\%$ and $32.8\,\%$ for the controls.

The full method also matches plain retrieval's 40-node recovery after returning only 24 nodes, or $40\,\%$ fewer returned nodes.

%% file: src/conclusions.tex
\section{Conclusion}

Our results show that graph structure can improve evidence retrieval without graph access at query time. Graph-modified embeddings guide the search for additional evidence, while the original embeddings maintain semantic relevance to the query. This supports developing graph-modified embeddings specifically for multi-stage retrieval and evaluating their contribution to the complete retrieval process.

Retrieval should adapt as evidence accumulates. The question and earlier results should guide when to use graph information, how to allocate the remaining search budget, and when enough evidence has been found to support an answer.

Future evaluation should test the approach on natural corpora, additional embedding models, and graphs with missing or incorrect relations. It should also establish whether smaller retrieved contexts offset the cost of additional searches. The broader goal is to design retrieval with the final answer in mind, using semantic similarity and graph structure together to provide the evidence needed within practical computational limits.

%% file: src/credit.tex
\section*{CRediT Author Statement}

\noindent\textbf{Federica Santi:}
Conceptualization, Data Curation, Formal Analysis, Investigation, Methodology, Software,
Validation, Visualization, Writing--Original Draft, Writing--Review \& Editing.

\noindent\textbf{Vasilije Markovi\'{c}:}
Conceptualization, Funding Acquisition, Project Administration, Resources, Supervision.

\noindent\textbf{Daniel Scott:}
Conceptualization, Data Curation, Formal Analysis, Investigation, Methodology, Software,
Validation, Visualization, Writing--Original Draft.

\noindent\textbf{Lazar Obradovi\'{c}:}
Conceptualization, Methodology,
Project Administration, Supervision, Writing--Review \& Editing.

%% file: src/appendix_overview.tex
	\section*{Appendix Overview}
	\begin{description}[leftmargin=0pt, labelsep=0.6em, itemsep=4pt]
		\item[\textnormal{\textbf{Appendix~\ref{app:data} --- Graph and dataset.}}]
		How the corpus was built: the generation procedure, the graph schema, the rendering of
		structured facts as text, the determinism guarantees, and the construction of the
		questions, answers, and gold-evidence sets.
		\item[\textnormal{\textbf{Appendix~\ref{app:retro} --- Retrofitting strategies.}}]
		The formulation of each embedding transformation, its parameter grid, and the
		implementation choices the main text states only in outline.
		\item[\textnormal{\textbf{Appendix~\ref{app:retr} --- Retrieval strategies.}}]
		The same for the retrievers: the stage schedules, their quotas, and the parameters that
		distinguish them.
		\item[\textnormal{\textbf{Appendix~\ref{app:perfamily} --- Per-family results.}}]
		Results for the six harder question families separately, whereas
		Section~\ref{sec:experiments} reports them pooled into groups.
	\end{description}

%% file: src/appendix_data.tex
	\section{Graph Construction and Evaluation Dataset}\label{app:data}
	\subsection{Graph Construction}
	The corpus is a fully synthetic professional network of fictional candidates, organisations, resumes, employer rosters, and textual evidence. Generation is deterministic under a fixed seed and does not involve a language model: every fact is sampled from predefined source pools and then rendered into text by fixed templates. Because the graph is built before any question exists, no answer information can enter its construction.

	\paragraph{Candidate identity.}
	A candidate's name is a first--last pair drawn from two flat pools of 512 and 458 entries. A \code{used\_names} set rejects any repeat and the draw is retried, so every person in the dataset has a distinct full name before the identity-collision checks performed at ingest. With roughly $2.3\times10^{5}$ available combinations and a few hundred candidates per graph, rejection is rare. Uniqueness is what allows the human-readable \code{entity:person:<Name>} identifier to serve as the canonical key: two candidates sharing a name would otherwise collapse into a single node and silently merge their evidence.

	\paragraph{What is generated.}
	Each candidate is assigned a primary sector sampled uniformly from eleven, which governs certifications and supplies the default source for employer and role attributes. A candidate receives one or two education records, between one and four jobs, zero to two projects per job, zero to three certifications, and zero to two endorsements. Education degrees are sampled with weights $50\,\%$ Bachelor's, $35\,\%$ Master's, and $15\,\%$ PhD, and universities come from a global pool that is not sector-scoped. Project descriptions reuse a subset of the skills attached to their job, so project evidence is correlated with the job rather than sampled independently.

	\paragraph{Controlled cross-sector bridges.}
	With probability $0.2$ a job's role attributes---title, skills, responsibilities, tenure, salary band, team size, and project name---are drawn from a different sector, selected through a fixed affinity map, while the employer remains tied to the candidate's primary sector. This prevents the graph from collapsing into isolated sector-specific clusters and produces known bridges between otherwise sector-oriented regions, such as technology roles inside healthcare organisations.

	\paragraph{Dual-perspective evidence.}
	Employment information is represented from two document perspectives. Resume-side chunks describe one person's education, jobs, projects, endorsements, and certifications. After all candidate histories are generated, a second pass groups jobs by employer: any employer with at least two employees receives a roster document and a roster chunk listing every employee and title, which mentions the organisation and each employee individually. The same employment fact therefore appears both in a person's job chunk and in an employer's roster chunk, so information about one entity is distributed across documents connected only through shared entities. This is what makes the corpus suitable for graph-retrieval experiments rather than reducible to independent flat documents.

	\paragraph{Schema.}
	The graph is a plain JSON dictionary with three node types and three edge types. Every node exposes an \code{embed\_field} naming the field embedded at ingest: \code{name} for entities and documents, \code{text} for chunks. The graph itself is therefore not embedded during generation.
	\begin{center}\small
		\begin{tabularx}{\textwidth}{@{}llY@{}}\toprule
			Category & Type & Semantics\\\midrule
			Node & \code{entity} & Person or organisation\\
			Node & \code{document} & Resume or roster\\
			Node & \code{chunk} & Textual evidence\\
			Edge & \code{part\_of} & Chunk $\rightarrow$ document\\
			Edge & \code{about} & Document $\rightarrow$ entity\\
			Edge & \code{mentions} & Chunk $\rightarrow$ entity\\\bottomrule
	\end{tabularx}\end{center}
	For each person the generator creates an \code{entity:person:<Name>} node and a \code{document:resume:<Name>} node joined by \code{about}. Every education, job, project, endorsement, and certification record becomes a chunk, linked to the resume by \code{part\_of} and to the person by \code{mentions}. Education chunks also mention a university entity and job and project chunks an employer entity; organisation nodes are created on first use and reused thereafter. Identifiers are deliberately human-readable---\code{entity:person:<Name>}, \code{chunk:job:<Name>:<i>}---and these are the canonical evaluation identifiers, preserved downstream in the vector store and in the result artefacts.

\begin{figure}[H]
	\centering
	\begin{tikzpicture}[
		x=1cm, y=1cm, font=\scriptsize,
		ch/.style={circle, draw=black!60, fill=black!6, minimum size=9pt, inner sep=0pt},
		en/.style={rectangle, draw=black!60, fill=black!6, minimum size=8pt, inner sep=0pt},
		dc/.style={rectangle, rounded corners=2pt, draw=black!60, fill=black!6,
			minimum size=8pt, inner sep=0pt},
		ed/.style={->, draw=black!45, line width=0.5pt, >=stealth},
		hl/.style={->, draw=colpick, line width=1pt, >=stealth},
		tx/.style={font=\scriptsize, inner sep=1pt},
		note/.style={font=\scriptsize\itshape, text=black!60},
		]
		\begin{scope}[shift={(0,0)}]
			\node[ch] (sc) at (0,0) {};      \node[tx, anchor=east] at (-0.2,0) {chunk};
			\node[dc] (sd) at (3.2,0) {};    \node[tx, anchor=south] at (3.2,0.18) {document};
			\node[en] (se) at (6.4,0) {};    \node[tx, anchor=west] at (6.6,0) {entity};
			\draw[ed] (sc) -- node[tx, above] {\code{part\_of}}
			node[tx, below, text=black!55] {1,171} (sd);
			\draw[ed] (sd) -- node[tx, above] {\code{about}}
			node[tx, below, text=black!55] {227} (se);
			\draw[ed] (sc) to[out=-28, in=-152]
			node[tx, below, pos=0.5] {\code{mentions}\ \textcolor{black!55}{2,267}} (se);
		\end{scope}
		\begin{scope}[shift={(0,-3.6)}]
			\node[ch] (j1) at (0, 0.75) {};
			\node[tx, anchor=east, align=right, text width=45mm] at (-0.25,0.75)
			{``Manish Henderson worked as a Social Services Manager at Eastgate Public Works
				Department\ldots''};
			\node[ch] (j2) at (0,-2.05) {};
			\node[tx, anchor=east, align=right, text width=45mm] at (-0.25,-2.05)
			{``Ana Ramos worked as a Litigation Support Specialist at Eastgate Public Works
				Department\ldots''};
			\node[dc] (d)  at (5.6, 1.75) {};
			\node[tx, anchor=west] at (5.85, 1.75) {\code{document:resume:Manish Henderson}};
			\node[en] (pe) at (5.6, 0.75) {};
			\node[tx, anchor=west] at (5.85, 0.75) {\code{entity:person:Manish Henderson}};
			\node[en] (o)  at (5.6,-1.05) {};
			\node[tx, anchor=west, align=left, text width=34mm] at (5.85,-1.05)
			{\code{entity:org:Eastgate Public Works Department}};
			\draw[ed] (j1) to[out=38, in=190]
			node[tx, pos=0.45, above, sloped] {\code{part\_of}} (d);
			\draw[ed] (d) -- node[tx, pos=0.5, right] {\code{about}} (pe);
			\draw[ed] (j1) -- node[tx, pos=0.5, above] {\code{mentions}} (pe);
			\draw[hl] (j1) -- node[tx, pos=0.42, below, sloped, text=black!70] {\code{mentions}} (o);
			\draw[hl] (j2) -- node[tx, pos=0.42, above, sloped, text=black!70] {\code{mentions}} (o);
			\node[note, anchor=north] at (2.2,-2.75)
			{the two coloured edges are one two-hop path: a chunk, a shared entity, another chunk};
		\end{scope}
	\end{tikzpicture}
	\caption{The graph schema and one instance from the validation corpus. Shape marks the node type:
		a circle is a chunk, a rounded square a document, a square an entity. Counts on the schema
		edges are those of the validation graph. In the instance, a resume document is \code{about} a
		person, its chunks are \code{part\_of} it, and each chunk \code{mentions} the entities named in
		its text. Because \code{mentions} always runs from a chunk to an entity, two hops lead from one
		chunk to another through an entity they share --- here two employees of the same organisation.}
	\label{fig:schema}
\end{figure}
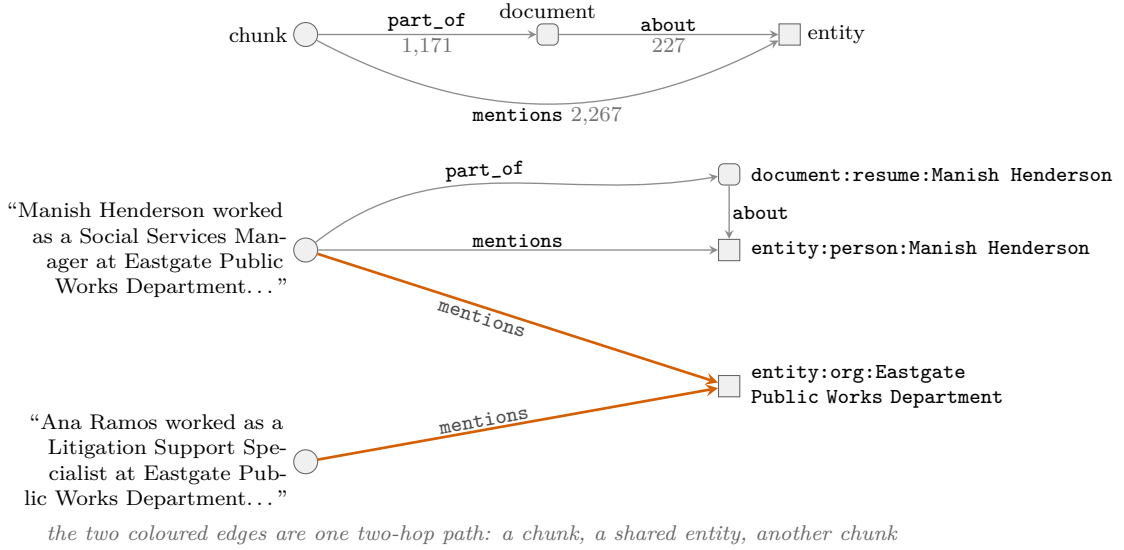

	The schema admits exactly three edge patterns and no others, so the graph is tripartite. One hop from a chunk reaches the entities it mentions and the document it belongs to; two hops return to other chunks through a shared entity. Figure~\ref{fig:schema} draws the schema and one instance from the validation corpus; the recurring chunk--entity--chunk route it shows is what the two-hop neighbour blend exploits.

	\paragraph{Structured facts and gold supervision.}
	Every generated fact is retained in two synchronised forms. Graph edges encode its structural relationships, and a human-readable rendering is stored in the corresponding chunk; the originating structured dictionary is preserved unchanged on that chunk as \code{source}. Text is therefore the representation exposed to embedding and retrieval, while \code{source} is the authoritative representation used to compute gold answers and supporting evidence. Question generation reads the structured copy, so exact gold evidence is obtained without an LLM reconstructing facts from prose.

	\paragraph{Determinism.}
	The generator seeds Python's global random module once with $\text{SEED}=42$ and proceeds strictly sequentially. The one hash-order-sensitive operation---the union of skills across a candidate's jobs, used to sample endorsements---is sorted before sampling, so process-dependent string hashing cannot affect the output. Under the same configuration and source pools the resulting \code{graph.json} is byte-identical across runs and machines, a property checked by \code{tests/test\_generation\_determinism.py} under different interpreter hash seeds.

	\paragraph{Development, validation, and test graphs.}
	A separate generation mode partitions the source pools themselves into disjoint development, validation, and test vocabularies using $\text{SPLIT\_SEED}=7$. Names, attributes, employers, and universities are split before the three graphs are generated, in the order development, validation, test, from a single seeded stream and a single shared \code{used\_names} set. No person name is therefore repeated across the three graphs, and the later graphs are not a mirrored copy of the earlier draws. Hyperparameter selection, generalisation checks, and final reporting do not share the same underlying vocabulary, and the corpora are node-disjoint by construction rather than by filtering.

	\paragraph{Validation corpus.}
	The validation graph used throughout this paper contains 1,759 nodes and 3,665 edges.
	\begin{center}\small
		\begin{tabularx}{\textwidth}{@{}lrY@{}}\toprule
			Component & Count & Composition\\\midrule
			Entities & 361 & 133 persons, 228 organisations\\
			Documents & 227 & 133 resumes, 94 rosters\\
			Chunks & 1,171 & 326 project, 319 job, 195 education, 133 endorsement, 104 certification, 94 roster\\\midrule
			\code{mentions} edges & 2,267 & chunk $\rightarrow$ entity\\
			\code{part\_of} edges & 1,171 & chunk $\rightarrow$ document\\
			\code{about} edges & 227 & document $\rightarrow$ entity\\\bottomrule
	\end{tabularx}\end{center}
	Each node carries an original embedding in $\mathbb{R}^{1536}$, and every modified collection preserves the same node identities, so a retrieved identifier means the same thing in every collection.

	\subsection{Questions and Gold Answers}
	Each question is paired with an exact answer and with the evidence required to derive it. Generation is deterministic under a fixed seed and uses no language model: Python functions traverse the graph, read the structured fact preserved in a chunk's \code{source} field, instantiate a fixed template, and record the nodes touched along the way. The generated chunk text is never parsed to recover an answer---text is what retrieval sees, while the structured record defines the ground truth.

	\paragraph{The stored record.}
	Every generator returns the triple $(q, C_q, a_q)$, where $q$ is the natural-language question, $C_q$ is the set of graph nodes touched while deriving the answer, and $a_q$ is the answer computed from structured facts. Before persistence $C_q$ is filtered to the evidence-bearing chunks, giving the gold-evidence set $G_q$, so each stored example is $(q, a_q^{\mathrm{gold}}, G_q^{\mathrm{gold}})$.

	\paragraph{Why only chunks count as targets.}
	Filtering keeps only nodes whose type is in \code{GOLDEN\_NODE\_TYPES = ("chunk",)}. Entities named in the question are query inputs, not retrieval targets: the retriever already has them. What it must recover is the text that carries the evidence. Rewarding a system for returning an entity the question itself names would measure nothing. Filtering happens at record-writing time only, and changes neither the question text, nor the computed answer, nor the sampling or reasoning depth.

	\paragraph{Question families.}
	Eleven families span basic, compositional, and multi-hop reasoning. Basic families retrieve or aggregate local attributes, compare candidates, or select extrema; compositional families chain two basic operations and union their evidence; multi-hop families traverse explicit person--organisation--university relations over one, two, or three hops, alternating between two predefined chains at each level. Because every traversal step is recorded, the benchmark keeps the whole supporting chain and not only the node holding the final answer.
	\begin{center}\small
		\begin{tabularx}{\textwidth}{@{}lYr@{}}\toprule
			Family & Main operation & Val.\\\midrule
			\code{collect\_local} & Read one attribute for one entity or entity--organisation pair & 90\\
			\code{collect\_global} & Aggregate one attribute across all people linked to an organisation & 90\\
			\code{comparison} & Compute a set intersection for two people & 90\\
			\code{extremize} & Select an argmax over job facts, optionally filtered and grouped & 77\\
			\code{extremize\_collect} & Find an extremal person, then retrieve a second attribute & 90\\
			\code{collect\_extremize} & Restrict to an organisation's people, then apply an extremum & 90\\
			\code{collect\_global\_collect} & Traverse through an organisation and retrieve a second attribute for each linked person & 90\\
			\code{multi\_collect} & Retrieve two attributes for one person in one question & 90\\
			\code{hop1} & One-hop organisation/person or university/person reasoning & 90\\
			\code{hop2} & Two-hop chains crossing entity types & 90\\
			\code{hop3} & Three-hop chains with aggregation or peer discovery & 90\\\midrule
			\multicolumn{2}{@{}l}{Validation set} & 977\\\bottomrule
	\end{tabularx}\end{center}

	One question of each family, taken from the validation set.
	\begin{center}\small
		\begin{tabularx}{\textwidth}{@{}lY@{}}\toprule
			Family & Example question\\\midrule
			\code{collect\_local} & ``What accreditations does Ethan Ito have?''\\
			\code{collect\_global} & ``What degrees has Union College awarded?''\\
			\code{comparison} & ``Were Naomi Das and Emily Hong ever at the same company?''\\
			\code{extremize} & ``Which candidate held a single position for the most years?''\\
			\code{extremize\_collect} & ``Which credentials has the most experienced grant management practitioner earned?''\\
			\code{collect\_extremize} & ``Who worked at Inkerman \& Price for the most years?''\\
			\code{collect\_global\_collect} & ``What certifications do employees of Juniper Health hold?''\\
			\code{multi\_collect} & ``What was Rajan Cook's tenure and what skills did they use at Xcel Freight Co?''\\
			\code{hop1} & ``Who attended University of Kentucky?''\\
			\code{hop2} & ``Where did people who worked at Jettison Labs study?''\\
			\code{hop3} & ``Which employer of Yale University graduates has the most total employees?''\\\bottomrule
	\end{tabularx}\end{center}

	\paragraph{Configuration-driven generation.}
	Most basic families are parameterised rather than implemented separately. A configuration names the relevant chunk and entity prefixes, whether an organisation is required, one or more surface templates, and an \code{answer\_fn} that extracts the target value from \code{source}. A generic routine resolves the linking chunk, applies the extraction function, and fills a template. Question semantics are therefore separated from traversal logic, and a new subtype costs a configuration entry rather than new generation code. Multi-hop generators are the exception: each chain is implemented directly, because successive hops cross different entity types.

	\paragraph{Sampling and budget.}
	Candidate tuples come from type-specific samplers---person--organisation, person pair, organisation, extremisation---each relying on seeded shuffling. Families with several subtypes cycle through their configurations so that consecutive questions alternate subtype until the budget is met or the candidate pool is exhausted. The nominal budget is divided evenly across the eleven families, so a request for 1,000 questions asks for 90 of each.

	Two families cannot fill that quota, and the reason is structural rather than incidental. \code{extremize} realises 77 on the validation graph, because an argmax needs a competition to resolve. \code{collect\_global\_collect} realises a single question, because some of its subtypes require one organisation to satisfy more than one semantic role at once, and few organisations in a single graph do. The validation set reported in this paper therefore takes the 888 questions realised on the validation graph and completes the under-filled family with 90 \code{collect\_global\_collect} questions generated on a second corpus from the same pipeline, for 977 in total.

	\paragraph{Gold-evidence set size.}
	Gold-evidence set size reflects the number of independent facts needed to establish an answer, not where the answer value happens to sit. A local extraction is usually one chunk: \code{q0000} asks how long a person worked at an organisation and its gold-evidence set is the single job chunk holding that record. Aggregation scales with the organisation: a question about tenure lengths across a company's staff carries one job chunk per employee. An argmax carries its competitors---\code{q0180} asks who is most experienced in a given skill and its gold-evidence set is the 15 job chunks that mention it, because the maximum cannot be verified without them. Compositional questions union the two steps' evidence, and three-hop questions produce the largest sets, up to 15 chunks spanning employer-to-employee, employee-to-university, and university-to-peer links.

	\paragraph{Determinism and materialisation.}
	Sampling and template choice draw from Python's global random generator, seeded once before generation. Each question receives a stable identifier of the form \code{q0000}, and \code{questions.jsonl} stores one deterministic JSON record per question: answer sets are serialised as sorted lists and gold-evidence node identifiers are sorted, so downstream evaluation joins on persistent identifiers rather than on text order or container iteration. Question embeddings are held in a separate vector collection, produced by the same embedding engine used for the graph-node collections, so query and document vectors share an embedding model by construction. Questions are stored as a plain model rather than as a graph data point, which keeps them out of any retriever's default collection scan.

%% file: src/appendix_retrofitting.tex
	\section{Retrofitting Strategies: Formulations and Implementation Details}\label{app:retro}

	The following sections consolidate the recorded implementation and parameter choices. Counts refer to generated collections, not all combinations with all retrievers. Unless otherwise stated, the blending coefficient retains the original representation with weight $\alpha$.

	\subsection{One-Hop Neighbour Blend}
	\code{hop1\_neighbor\_blend} averages a subset of the node's direct neighbours and mixes that mean with the original vector.
	\begin{center}\small
		\begin{tabularx}{\textwidth}{@{}lYl@{}}\toprule
			Parameter & Meaning & Values\\\midrule
			\code{neighbor\_pool\_share} & Fraction of available direct neighbours used & 0.2, 0.5, 1.0\\
			\code{alpha} & Weight retained from the original vector & 0.0, 0.2, 0.5\\
			\code{neighbor\_pick} & Select least- or most-similar eligible neighbours & \code{least\_similar}, \code{most\_similar}\\
			\code{min\_neighbors} & Minimum selected count when the fraction is small & 2 (\fixed)\\\bottomrule
	\end{tabularx}\end{center}
	The share is degree-relative: at 20\,\%, a node with 60 neighbours uses 12, while a node with five uses the minimum of two. A full Cartesian grid would contain 18 combinations. At share 1.0, however, both selection orders include the entire pool and produce byte-identical outputs; only one copy is retained, leaving 15 collections.

	\subsection{Two-Hop Neighbour Blend}
	\code{hop2\_only\_neighbor\_blend} uses nodes at graph distance exactly two, excluding the source and its direct neighbours. It uses the same share, $\alpha$, selection, and minimum-neighbour grid as the one-hop neighbour blend. The additional parameter \code{neighbor\_vectors} is fixed to \code{original}: the method always averages the original embeddings of two-hop nodes. Blending already-modified neighbour vectors was deliberately not implemented because its interaction with share-based selection was not specified.

	Along chunk--entity--chunk paths, two hops often map a chunk to another chunk through a shared entity. The method therefore captures other texts that discuss the same entities rather than the entities themselves.

	\subsection{Convex Neighbour Pull}
	This anchored neighbour-attraction formulation is related to retrofitting~\cite{faruqui}, applied here to corpus-node embeddings.
	\code{convex\_neighbor\_pull} jointly adjusts node embeddings by balancing fidelity to their original textual representations against agreement between graph neighbours. A useful physical interpretation is a system of springs: each node is attached to its original position and to every adjacent node. The modified embeddings are the equilibrium of this coupled system.

	\paragraph{Objective and uniqueness.}
	Let $\mathbf{x}_i$ be the original embedding and $\mathbf{z}_i$ the unknown modified embedding. Let $E$ contain unordered edge pairs, each counted once, and let $N(i)$ be the corresponding neighbour set. The objective is
	\begin{equation}
		J(\mathbf Z)=a\sum_i\|\mathbf z_i-\mathbf x_i\|_2^2
		+b\sum_{(i,j)\in E}\|\mathbf z_i-\mathbf z_j\|_2^2.
		\label{eq:convex-objective}
	\end{equation}
	Here $a=\code{anchor\_weight}>0$ and $b=\code{edge\_weight}\geq0$. We use $a$ rather than $\alpha$ to distinguish this penalty weight from the blending coefficient used by other transformations. Squared Euclidean distances yield a differentiable quadratic objective with linear stationarity equations. The positive anchor term makes the objective strictly convex, so its minimiser is unique. The implementation rejects nonpositive anchor weights. With $a=0$ and $b>0$, any embedding configuration constant within each connected component has zero energy; the minimiser is no longer unique.

	\paragraph{Closed-form coordinate update.}
	Setting the derivative with respect to one node to zero gives
	\begin{equation}
		a(\mathbf z_i-\mathbf x_i)+b\sum_{j\in N(i)}(\mathbf z_i-\mathbf z_j)=0,
	\end{equation}
	and hence the weighted-average update
	\begin{equation}
		\mathbf z_i\leftarrow
		\frac{a\mathbf x_i+b\sum_{j\in N(i)}\mathbf z_j}{a+b|N(i)|}.
		\label{eq:convex-update}
	\end{equation}
	The original vector has weight $a$, while each neighbour has weight $b$. Thus, total graph influence grows with degree: the denominator is $a+3b$ for three neighbours and $a+60b$ for sixty. An isolated node remains at its original vector. Although each coordinate update has a closed form, the global solution requires solving the coupled system.

	\paragraph{Gauss--Seidel solver.}
	The solver initialises $\mathbf Z$ with the original embeddings and visits nodes in sorted identifier order. Each update is written immediately, so later nodes in the same sweep use the newest available neighbour vectors. This is Gauss--Seidel iteration, rather than a synchronous Jacobi update based entirely on the previous sweep. It requires no learning rate or gradient clipping, although the objective weights and stopping controls remain configuration choices. For this anchored quadratic, convergence is to the same unique minimiser; finite-sweep results and convergence speed can depend on update order. Sorted identifiers ensure a reproducible order.

	\paragraph{Stopping rule.}
	During sweep $t$, the solver records the largest Euclidean displacement over all node updates,
	\begin{equation}
		\delta_t=\max_i\|\mathbf z_i^{\mathrm{after\ update}}-
		\mathbf z_i^{\mathrm{before\ update}}\|_2.
	\end{equation}
	The displacement is measured before overwriting each vector. Convergence is declared when $\delta_t<10^{-6}$; the criterion uses the maximum rather than the mean, so every node must satisfy the tolerance. On the reported development graph of 1,811 nodes, convergence took approximately forty sweeps, with a final maximum displacement of $9\times10^{-7}$. A safety cap of 200 sweeps returns the current configuration even if convergence has not been reached; the manifest records the sweep count and final displacement so that incomplete convergence remains visible. A separate early-stop setting terminates after a prescribed number of sweeps.
	\begin{center}\small
		\begin{tabularx}{\textwidth}{@{}lYl@{}}\toprule
			Parameter & Meaning & Values\\\midrule
			\code{anchor\_weight} & Strength of the original-vector fidelity term & 1.0 (\fixed)\\
			\code{edge\_weight} & Strength of attraction along graph edges & 0.5, 1.0\\
			\code{convergence\_tol} & Stop when vector movement falls below this threshold & $10^{-6}$ (\fixed)\\
			\code{max\_sweeps} & Safety cap on full update sweeps & 200 (\fixed)\\
			\code{stop\_after\_sweeps} & Optional forced early stop & None, 1, 2, 3\\\bottomrule
	\end{tabularx}\end{center}
	\paragraph{Grid construction.}
	Multiplying both weights by the same positive constant rescales the objective without changing its minimiser or update rule. Only their ratio matters, so $a=1$ fixes the reference scale. The grid contains two runs to convergence, with $b\in\{0.5,1.0\}$, and three early-stop probes at $b=1.0$ after one, two, or three sweeps. These are five configurations, not the full Cartesian product of the listed values. \code{stop\_after\_sweeps=None} uses the convergence criterion subject to the safety cap.

	\paragraph{Early-stopping probe.}
	The probe tests whether stopping short of the optimum preserves useful structure that full smoothing might remove. In the reported comparison, F1 increases across the tested sweep counts and is highest at convergence:
	\begin{center}\small
		\begin{tabular}{@{}lrrrr@{}}\toprule
			& 1 sweep & 2 sweeps & 3 sweeps & Converged\\\midrule
			F1 & 0.0933 & 0.1051 & 0.1120 & 0.1178\\\bottomrule
	\end{tabular}\end{center}
	These measurements provide no evidence of a benefit from the tested early stops. They do not establish monotonic retrieval improvement at every unmeasured iteration.

	\paragraph{Retrieval behaviour.}
	The separately reported usage comparison gives F1 of 0.1202 for standalone retrieval and 0.1510 when this representation is used in the single requery schedule. A random-walk endpoint configuration scores 0.1534, a numerical difference of 0.0024, without an iterative convergence phase. This is a comparison of the supplied scores, not a claim of statistical equivalence. It illustrates that minimising the graph-smoothing objective and maximising retrieval quality are different goals. The reported ranks (491 and 25 in an approximately 600-entry comparison) are omitted because the exact ranking population is not specified relative to the configurations documented here.

	\subsection{Random-Walk Endpoint Blend}
	For each node $i$ whose embedding is being modified, \code{walk\_endpoint} launches 100 random walks starting from that same node $i$. It forms the graph-context vector for $i$ as a frequency-weighted mean of the endpoint embeddings; intermediate paths are discarded.
	\begin{center}\small
		\begin{tabularx}{\textwidth}{@{}lYl@{}}\toprule
			Parameter & Meaning & Values\\\midrule
			\code{walk\_steps} & Number of graph transitions per walk & 2, 3\\
			\code{alpha} & Original-vector retention & 0.0, 0.5, 0.8\\
			\code{walks\_per\_node} & Walks launched from each node & 100 (\fixed)\\
			\code{walk\_seed} & Reproducibility seed & 42 (\fixed)\\\bottomrule
	\end{tabularx}\end{center}
	Along alternating chunk--entity paths, two transitions can return to a chunk and three can end on an entity. Other node and edge types may change this pattern. Returns to the starting node are possible unless excluded by the transition rule; that rule remains to be verified.

	\subsection{Louvain Cluster Blend}
	Community detection uses the Louvain family introduced by Blondel et al.~\cite{blondel}; the subsequent embedding blend is the transformation evaluated here.
	\code{louvain\_cluster} detects graph communities and moves each node towards the centroid of its community.
	\begin{center}\small
		\begin{tabularx}{\textwidth}{@{}lYl@{}}\toprule
			Parameter & Meaning & Values\\\midrule
			\code{resolution} & Controls community granularity; larger values produce smaller communities & 0.5, 1.0, 2.0\\
			\code{alpha} & Original-vector retention & 0.0, 0.5, 0.8\\
			\code{louvain\_seed} & Reproducible partitioning seed & 42 (\fixed)\\\bottomrule
	\end{tabularx}\end{center}
	At resolutions 0.5, 1.0, and 2.0, the graph contains respectively 20, 34, and 50 communities, with median sizes 85, 44, and 31. When $\alpha=0$, every member is mapped to its frozen original-space community centroid, eliminating within-community distinctions.

	\subsection{Shared-Neighbour Pull}
	\code{shared\_neighbor\_pull} adds an attraction term between nodes that are not directly connected but share graph neighbours.
	\begin{center}\small
		\begin{tabularx}{\textwidth}{@{}lYl@{}}\toprule
			Parameter & Meaning & Values\\\midrule
			\code{shared\_pair\_weight} & Strength of attraction between common-neighbour pairs & 1.0, 3.0, 9.0\\
			\code{stop\_after\_sweeps} & Optional forced early stop & None, 1, 2, 3\\
			\code{anchor\_weight} & Original-vector fidelity term & 1.0 (\fixed)\\
			\code{edge\_weight} & Direct-edge attraction term & 1.0 (\fixed)\\
			\code{convergence\_tol} & Convergence threshold & $10^{-6}$ (\fixed)\\
			\code{max\_sweeps} & Safety cap & 200 (\fixed)\\\bottomrule
	\end{tabularx}\end{center}
	The coefficient values are 1, 3, and 9 with direct-edge weight fixed to 1. Their aggregate influence also depends on the number and weighting of eligible shared-neighbour pairs. Three configurations converge fully; three early-stop probes use the weight-3 setting. Fixing the edge weight to one makes \code{convex\_neighbor\_pull\_\_e1p0} an exact zero-shared-pair control under the same solver.

	\subsection{Gradient-Based Neighbour Pull}
	\code{gradient\_neighbor\_pull} optimises embeddings so that vector similarity follows graph-hop distance.
	\begin{center}\small
		\begin{tabularx}{\textwidth}{@{}lYl@{}}\toprule
			Parameter & Meaning & Values\\\midrule
			\code{hop\_target\_curve} & Desired similarity decay over hop distance & None, \code{linear\_steep}, \code{hyperbolic\_sech}\\
			\code{step\_size} & Magnitude of each optimisation update & 0.25, 0.5, 1.0\\
			\code{max\_shift} & Maximum displacement as a fraction of vector norm & 0.5, 1.0, 2.0\\
			\code{steps} & Number of optimisation passes & 30 (\fixed)\\
			\code{patch\_hops} & Radius of overlapping graph patches & 2 (\fixed)\\
			\code{hop\_target\_max\_hop} & Hop distance beyond which target similarity is zero & 4 (\fixed)\\
			\code{patch\_seed} & Reproducibility seed & 42 (\fixed)\\\bottomrule
	\end{tabularx}\end{center}
	The grid is intentionally sparse: one default configuration uses no explicit curve, step 0.5, and shift 1.0; the two named curves are crossed with two step sizes and two shift limits, producing eight additional configurations. This is the only transformation with a learning-rate-like step and an explicit displacement cap.

	\subsection{Residual Representations}
	A residual has no independent hyperparameters. For each of the 22 principal swept collections, it subtracts the normalised original vector from the normalised modified vector. The sole parameter, \code{modified\_collection}, selects the source collection.

%% file: src/appendix_retrieval.tex
	\section{Retrieval Strategies: Formulations and Implementation Details}\label{app:retr}
	Every retriever below is evaluated under the invariant final budget $B=40$, and each is crossed with the derived collections. Two accessory grids re-run part of the shortlist at half the budget and over a chunk-only search domain.

	\subsection{Graph-Agnostic Embedding Retrieval and One Modified Collection Alone}
	\code{MergedSearchRetriever} performs standard nearest-neighbour retrieval over a chosen set of stored collections.
	\begin{center}\small
		\begin{tabularx}{\textwidth}{@{}lYl@{}}\toprule
			Parameter & Meaning & Values\\\midrule
			\code{collections} & The index set searched & the three originals jointly, or one derived collection\\\bottomrule
	\end{tabularx}\end{center}
	One run searches the three original collections jointly and defines \code{plain\_baseline}; the others search one derived collection directly.

	\subsection{Alternating Budget Split}
	\code{BudgetSplitRetriever} searches the original and the modified space independently, then allocates the final positions between the two rankings.
	\begin{center}\small
		\begin{tabularx}{\textwidth}{@{}lYl@{}}\toprule
			Parameter & Meaning & Values\\\midrule
			\code{beta} & Share of the final budget given to the modified space & 0.5, 0.2\\\bottomrule
	\end{tabularx}\end{center}
	At $\beta=0.5$, 20 positions go to each space; at $\beta=0.2$, 8 go to text and 32 to the modified space. Thirty hop-derived collections are crossed with both values, and the fixed-count collection is evaluated once at $\beta=0.5$.

	\subsection{Reranking}
	\code{RerankRetriever} retrieves a deep pool from the original space and rescores it with a graph-derived signal, scoring each candidate as
	\begin{equation}
		s_i=(1-\texttt{mix})\,\mathrm{sim}(\mathbf q,\mathbf x_i)+\texttt{mix}\,\mathrm{sim}(\mathbf m,\mathbf z_i).
	\end{equation}
	\begin{center}\small
		\begin{tabularx}{\textwidth}{@{}lYl@{}}\toprule
			Parameter & Meaning & Values\\\midrule
			\code{mix} & Weight of the graph-derived score against the original-space score & 0.3, 0.5, 0.7\\
			\code{RERANK\_OVERSAMPLE} & Candidate pool as a multiple of the budget & 4 (\fixed)\\\bottomrule
	\end{tabularx}\end{center}
	The 22 modified collections are crossed with the three mix values. Mix zero is excluded because it exactly reproduces the text-only ranking. The oversampling factor fixes the pool at $4B=160$, and the seed count is derived as $B/2=20$; neither is treated as a tunable parameter.

	\subsection{Two-Stage Fused Retrieval and Requery Schedules}
	\code{MultiStageRetriever} fills the budget sequentially, one stage at a time. Seven stage schedules are evaluated:
	\begin{center}\small
		\begin{tabularx}{\textwidth}{@{}clY@{}}\toprule
			Quotas & Stage ranks & Interpretation\\\midrule
			16/12/12 & original, fused, shifted & Single requery: expansion followed by query-shifted text retrieval\\
			6/6/14/14 & original, fused, shifted, shifted & Chained requery: a short opening, then two successive shifts\\
			20/20 & original, fused & Two-stage fused retrieval: text seeds followed by modified-space expansion\\
			16/8/8/8 & original, fused, shifted, shifted & The same chaining from a wide opening\\
			6/12/22 & original, fused, shifted & Single requery from a short opening\\
			6/6/14/14 & original, fused, fused, fused & The chained quotas with every later stage fused instead of shifted\\
			20/20 & original, shifted & Graph-free control: the shift alone, with no fused stage\\\bottomrule
	\end{tabularx}\end{center}
	\begin{center}\small
		\begin{tabularx}{\textwidth}{@{}lYl@{}}\toprule
			Parameter & Meaning & Values\\\midrule
			\code{mix} & Weight of the modified-space score in a fused stage & 0.3, 0.5, 0.7\\
			\code{query\_shift} & Weight of the previous stage's mean in the shifted query & 0.3, 0.5 (\code{bridge} only)\\\bottomrule
	\end{tabularx}\end{center}
	The shifted query is
	\begin{equation}
		\mathbf q'=(1-\texttt{query\_shift})\mathbf q+\texttt{query\_shift}\,\bar{\mathbf x}_{\mathrm{previous}}.
	\end{equation}
	The coefficient is a nominal weight: the norm of the mean determines its effective contribution, so a coefficient of 0.5 need not produce equal directional influence. The chained schedule carries two parameters the others do not: beyond the mix and query-shift grids it also varies which retrieved nodes a shifted stage anchors on and how far back it looks. The modified-space input ranges over 44 representations: 22 graph-modified collections and their 22 residuals.

	\subsection{Output Ordering, Overlap, and Computational Scope}
	Single requery (\code{bridge}) is formulated in Section~\ref{retr-sect}. Its final output concatenates the three stage lists; it is not globally reranked. Stage-level exclusion uses node identity. Candidate-depth and refill rules must ensure that an implementation actually returns 40 distinct nodes, including when independent searches overlap. For the alternating budget split, the quota and overlap-resolution convention must be recovered from the implementation; taking two top lists alone does not guarantee 40 distinct nodes.

	The recorded search-call counts are one for plain retrieval, two for budget splitting, one for reranking, three for two-stage fused retrieval, four for single requery, and five for the four-stage chained schedules. They count vector-search operations rather than equal units of work. Reranking evaluates 160 candidates; modified collections require preprocessing and storage. No latency or hardware measurements are available.

%% file: src/appendix_per_family.tex
	\section{Per-Family Results}\label{app:perfamily}
	Section~\ref{sec:experiments} reports the question families pooled into groups. The table below gives each of the six harder families separately, for the retrieval strategies described in Section~\ref{retr-sect}. The five single-fact families are omitted: the baseline already recovers between $96.4\,\%$ and $100\,\%$ of their evidence, so no strategy can separate itself on them except by degrading them. Each column uses the representation the main text pairs with that schedule: the hop-2 neighbour blend for the alternating budget split and for two-stage fused retrieval, the walk-endpoint blend for both requery schedules. The baseline column is plain vector retrieval on the original collections.

	\begin{center}\small
		\begin{tabularx}{\textwidth}{@{}lYrrrrr@{}}\toprule
			& & \multicolumn{5}{c}{Gold chunks found (\%)}\\\cmidrule(l){3-7}
			Group & Family & Base & Split & Fused & Single & Chained\\
			& & & \multicolumn{2}{c}{\footnotesize hop-2} & \multicolumn{2}{c}{\footnotesize walk}\\\midrule
			\multirow{3}{*}{Extremisation}
			& \code{collect\_extremize} & 72.4 & 71.6 & 86.6 & 87.5 & \textbf{88.1}\\
			& \code{extremize} & 61.9 & 42.5 & 54.5 & 68.4 & \textbf{75.0}\\
			& \code{extremize\_collect} & 44.7 & 32.4 & 40.6 & 59.7 & \textbf{61.7}\\\midrule
			\multirow{3}{*}{Relational and collection}
			& \code{collect\_global\_collect} & 66.4 & \textbf{85.7} & 80.3 & 70.0 & 71.6\\
			& \code{hop2} & 42.3 & \textbf{64.7} & 48.5 & 46.7 & 50.0\\
			& \code{hop3} & 20.1 & \textbf{31.9} & 22.7 & 20.6 & 20.2\\\bottomrule
	\end{tabularx}\end{center}

	Two patterns are visible here that the group-level reporting averages away.

	The first is that the two leading strategies are complementary rather than ranked. On the three extremisation families the chained requery is best everywhere, by 13.2 to 17.0 points over the baseline; on the three relational and collection families the order reverses and the split is best on all three, by 11.8 to 22.4 points. Which strategy helps therefore depends on the operation a question requires --- covering a comparison set, or reaching along a relational chain --- and not on a single ordering of methods.

	The two failures are not symmetric, and the difference matters for anyone who must choose one schedule for a mixed workload. Outside its own group the split is actively harmful: it falls below the baseline on all three extremisation families, by $0.8$, $12.3$ and $19.3$ points. The chained requery outside its own group is merely modest: it stays above the baseline on all three relational and collection families, by $5.2$, $7.6$ and $0.1$ points, the last of which is indistinguishable from zero. Choosing the split therefore costs real evidence on the questions it does not suit, whereas choosing the chained schedule costs almost nothing.

	The second is that \code{hop3} resists every strategy. Its baseline is the lowest in the benchmark at $20.1\,\%$ and the best column improves it by $11.8$ points, while both requery schedules leave it where it was. The limit is not the budget: 40 retrieved nodes are ample for a median gold-evidence set of twelve chunks, and doubling the budget from 20 to 40 nodes gains the baseline only $1.6$ points, because the extra nodes repeat content the ranking has already exhausted.

%% file: main.bbl
\begin{thebibliography}{99}
		\bibitem{lewis2020rag}
		Lewis, P., Perez, E., Piktus, A., Petroni, F., Karpukhin, V., Goyal, N., K{\"u}ttler, H., Lewis, M., Yih, W.-T., Rockt{\"a}schel, T., Riedel, S., \& Kiela, D. (2020). Retrieval-augmented generation for knowledge-intensive NLP tasks. In Advances in Neural Information Processing Systems, vol. 33 (pp. 9459-9474). \url{https://arxiv.org/abs/2005.11401}.
		\bibitem{karpukhin2020dpr}
		Karpukhin, V., O{\u{g}}uz, B., Min, S., Lewis, P., Wu, L., Edunov, S., Chen, D., \& Yih, W.-T. (2020). Dense passage retrieval for open-domain question answering. In Proceedings of the 2020 Conference on Empirical Methods in Natural Language Processing (pp. 6769-6781). Association for Computational Linguistics. \url{https://doi.org/10.18653/v1/2020.emnlp-main.550}.
		\bibitem{faruqui}
		Faruqui, M., Dodge, J., Jauhar, S. K., Dyer, C., Hovy, E., \& Smith, N. A. (2015). Retrofitting word vectors to semantic lexicons. In Proceedings of the 2015 Conference of the North American Chapter of the Association for Computational Linguistics: Human Language Technologies (pp. 1606-1615). Association for Computational Linguistics. \url{https://doi.org/10.3115/v1/N15-1184}.
		\bibitem{mrksic}
		Mrk\v{s}i\'{c}, N., \'{O} S\'{e}aghdha, D., Thomson, B., Ga\v{s}i\'{c}, M., Rojas-Barahona, L. M., Su, P.-H., Vandyke, D., Wen, T.-H., \& Young, S. (2016). Counter-fitting word vectors to linguistic constraints. In Proceedings of the 2016 Conference of the North American Chapter of the Association for Computational Linguistics: Human Language Technologies (pp. 142-148). Association for Computational Linguistics. \url{https://doi.org/10.18653/v1/N16-1018}.
		\bibitem{lengerich}
		Lengerich, B., Maas, A., \& Potts, C. (2018). Retrofitting distributional embeddings to knowledge graphs with functional relations. In Proceedings of the 27th International Conference on Computational Linguistics (pp. 2423-2436). Association for Computational Linguistics. \url{https://aclanthology.org/C18-1205/}.
		\bibitem{graftnet}
		Sun, H., Dhingra, B., Zaheer, M., Mazaitis, K.,
		Salakhutdinov, R., \& Cohen, W. W. (2018).
		Open domain question answering using early fusion of knowledge bases
		and text.
		In \emph{Proceedings of EMNLP 2018} (pp. 4231--4242).
		\url{https://doi.org/10.18653/v1/D18-1455}.
		\bibitem{pullnet}
		Sun, H., Bedrax-Weiss, T., \& Cohen, W. W. (2019).
		PullNet: Open domain question answering with iterative retrieval on
		knowledge bases and text.
		In \emph{Proceedings of EMNLP-IJCNLP 2019} (pp. 2380--2390).
		\url{https://doi.org/10.18653/v1/D19-1242}.
		\bibitem{nprf}
		Li, C., Sun, Y., He, B., Wang, L., Hui, K., Yates, A., Sun, L.,
		\& Xu, J. (2018).
		NPRF: A neural pseudo relevance feedback framework for ad-hoc
		information retrieval.
		In \emph{Proceedings of EMNLP 2018} (pp. 4482--4491).
		\url{https://doi.org/10.18653/v1/D18-1478}.
		\bibitem{colbertprf}
		Wang, X., Macdonald, C., Tonellotto, N., \& Ounis, I. (2021).
		Pseudo-relevance feedback for multiple representation dense retrieval.
		In \emph{Proceedings of ICTIR 2021} (pp. 297--306).
		\url{https://doi.org/10.1145/3471158.3472250}.
		\bibitem{yuprf}
		Yu, H., Xiong, C., \& Callan, J. (2021).
		Improving query representations for dense retrieval with
		pseudo-relevance feedback.
		In \emph{Proceedings of the 30th ACM International Conference on
			Information and Knowledge Management (CIKM '21)}
		(pp. 3598--3602).
		\url{https://doi.org/10.1145/3459637.3482124}.
		\bibitem{blondel}
		Blondel, V.~D., Guillaume, J.-L., Lambiotte, R., and Lefebvre, E.
		Fast unfolding of communities in large networks.
		\textit{Journal of Statistical Mechanics: Theory and Experiment},
		2008(10), 2008.
		\url{https://doi.org/10.1088/1742-5468/2008/10/P10008}.
		\bibitem{cattaneo}
		Cattaneo, A., Luschi, C., \& Justus, D. (2025). Ground-Truth Subgraphs for Better Training and Evaluation of Knowledge Graph Augmented LLMs. arXiv preprint arXiv:2511.04473. \url{https://doi.org/10.48550/arXiv.2511.04473}.
		\bibitem{cai}
		Cai, D., Li, X., Ho, J. C.-S., Bing, L., \& Lam, W. (2022). Retrofitting multilingual sentence embeddings with Abstract Meaning Representation. In Proceedings of the 2022 Conference on Empirical Methods in Natural Language Processing (pp. 6456-6472). Association for Computational Linguistics. \url{https://doi.org/10.18653/v1/2022.emnlp-main.433}.
		\bibitem{mizuki}
		Mizuki, S., \& Okazaki, N. (2023). Semantic specialization for knowledge-based word sense disambiguation. In Proceedings of the 17th Conference of the European Chapter of the Association for Computational Linguistics (pp. 3457-3470). Association for Computational Linguistics. \url{https://doi.org/10.18653/v1/2023.eacl-main.251}.
		\bibitem{xie}
		Xie, R., Liu, Z., Jia, J., Luan, H., \& Sun, M. (2016). Representation learning of knowledge graphs with entity descriptions. Proceedings of the AAAI Conference on Artificial Intelligence, 30(1). \url{https://doi.org/10.1609/aaai.v30i1.10329}.
		\bibitem{wang}
		Wang, L., Zhao, W., Wei, Z., \& Liu, J. (2022). SimKGC: Simple contrastive knowledge graph completion with pre-trained language models. In Proceedings of the 60th Annual Meeting of the Association for Computational Linguistics (Volume 1: Long Papers) (pp. 4281-4294). Association for Computational Linguistics. \url{https://doi.org/10.18653/v1/2022.acl-long.295}.
		\bibitem{li}
		Li, Q., Zhong, Y., \& Qin, Y. (2024). MoCoKGC: Momentum contrast entity encoding for knowledge graph completion. In Proceedings of the 2024 Conference on Empirical Methods in Natural Language Processing (pp. 14940-14952). Association for Computational Linguistics. \url{https://doi.org/10.18653/v1/2024.emnlp-main.832}.
		\bibitem{qiu}
		Qiu, C., Qian, P., Wang, C., Yao, J., Liu, L., Fang, W., \& Ng, E.-Y.-K. (2024). Joint pre-encoding representation and structure embedding for efficient and low-resource knowledge graph completion. In Proceedings of the 2024 Conference on Empirical Methods in Natural Language Processing (pp. 15257-15269). Association for Computational Linguistics. \url{https://doi.org/10.18653/v1/2024.emnlp-main.851}.
		\bibitem{papaluca}
		Papaluca, A., Krefl, D., Lensky, A., \& Suominen, H. (2025). Contrastive language-entity pre-training for richer knowledge graph embedding. In C. Wallraven, C.-L. Liu, \& A. Ross (Eds.), Pattern Recognition and Artificial Intelligence - ICPRAI 2024, Lecture Notes in Computer Science, vol. 14892 (pp. 233-246). Springer. \url{https://doi.org/10.1007/978-981-97-8702-9_16}.
		\bibitem{perozzi2014deepwalk}
		Perozzi, B., Al-Rfou, R., \& Skiena, S. (2014). DeepWalk: Online learning of social representations. In Proceedings of the 20th ACM SIGKDD International Conference on Knowledge Discovery and Data Mining (pp. 701-710). Association for Computing Machinery. \url{https://doi.org/10.1145/2623330.2623732}.
		\bibitem{grover2016node2vec}
		Grover, A., \& Leskovec, J. (2016). node2vec: Scalable feature learning for networks. In Proceedings of the 22nd ACM SIGKDD International Conference on Knowledge Discovery and Data Mining (pp. 855-864). Association for Computing Machinery. \url{https://doi.org/10.1145/2939672.2939754}.
		\bibitem{wu2019sgc}
		Wu, F., Souza, A., Zhang, T., Fifty, C., Yu, T., \& Weinberger, K. Q. (2019). Simplifying graph convolutional networks. In Proceedings of the 36th International Conference on Machine Learning, Proceedings of Machine Learning Research, vol. 97 (pp. 6861-6871). PMLR. \url{https://proceedings.mlr.press/v97/wu19e.html}.
		\bibitem{xiong2021multihop}
		Xiong, W., Li, X. L., Iyer, S., Du, J., Lewis, P., Wang, W. Y., Mehdad, Y., Yih, W.-T., Riedel, S., Kiela, D., \& O{\u{g}}uz, B. (2021). Answering complex open-domain questions with multi-hop dense retrieval. In International Conference on Learning Representations. \url{https://arxiv.org/abs/2009.12756}.
		\bibitem{trivedi2023ircot}
		Trivedi, H., Balasubramanian, N., Khot, T., \& Sabharwal, A. (2023). Interleaving retrieval with chain-of-thought reasoning for knowledge-intensive multi-step questions. In Proceedings of the 61st Annual Meeting of the Association for Computational Linguistics (Volume 1: Long Papers) (pp. 10014-10037). Association for Computational Linguistics. \url{https://doi.org/10.18653/v1/2023.acl-long.557}.
		\bibitem{shao2023iterretgen}
		Shao, Z., Gong, Y., Shen, Y., Huang, M., Duan, N., \& Chen, W. (2023). Enhancing retrieval-augmented large language models with iterative retrieval-generation synergy. In Findings of the Association for Computational Linguistics: EMNLP 2023 (pp. 9248-9274). Association for Computational Linguistics. \url{https://doi.org/10.18653/v1/2023.findings-emnlp.620}.
		\bibitem{rocchio1971relevance}
		Rocchio, J. J., Jr. (1971). Relevance feedback in information retrieval. In G. Salton (Ed.), The SMART Retrieval System: Experiments in Automatic Document Processing. Prentice-Hall.
		\bibitem{yang2018hotpotqa}
		Yang, Z., Qi, P., Zhang, S., Bengio, Y., Cohen, W. W., Salakhutdinov, R., \& Manning, C. D. (2018). HotpotQA: A dataset for diverse, explainable multi-hop question answering. In Proceedings of the 2018 Conference on Empirical Methods in Natural Language Processing (pp. 2369-2380). Association for Computational Linguistics. \url{https://doi.org/10.18653/v1/D18-1259}.
		\bibitem{trivedi2022musique}
		Trivedi, H., Balasubramanian, N., Khot, T., \& Sabharwal, A. (2022). MuSiQue: Multihop questions via single-hop question composition. Transactions of the Association for Computational Linguistics, 10, 539-554. \url{https://doi.org/10.1162/tacl_a_00475}.
		\bibitem{gutierrez2024hipporag}
		Jim{\'e}nez Guti{\'e}rrez, B., Shu, Y., Gu, Y., Yasunaga, M., \& Su, Y. (2024). HippoRAG: Neurobiologically inspired long-term memory for large language models. In Advances in Neural Information Processing Systems. \url{https://arxiv.org/abs/2405.14831}.
		\bibitem{edge2024graphrag}
		Edge, D., Trinh, H., Cheng, N., Bradley, J., Chao, A., Mody, A., Truitt, S., Metropolitansky, D., Ness, R. O., \& Larson, J. (2024). From local to global: A graph RAG approach to query-focused summarization. arXiv preprint arXiv:2404.16130. \url{https://doi.org/10.48550/arXiv.2404.16130}.
	\end{thebibliography}
